\documentclass[reqno]{amsart}

\usepackage{color}
\usepackage{geometry}
\usepackage{graphicx}
\usepackage[section]{placeins}
\usepackage{hyperref}
\usepackage{stmaryrd}
\usepackage{amsmath,amssymb,amsfonts,amsthm,textcomp}
\usepackage[foot]{amsaddr}
\usepackage{mathtools}
\usepackage{subfig}
\usepackage[numbers,sort&compress,square]{natbib}
\graphicspath{{./figures/}}
\usepackage[export]{adjustbox}

\usepackage{mathrsfs}
\usepackage{siunitx}
\usepackage{chemformula}

\newcommand{\bfe}{\boldsymbol{e}}
\newcommand{\bff}{\boldsymbol{f}}

\newcommand{\bfj}{\boldsymbol{\jmath}}

\newcommand{\bfn}{\boldsymbol{n}}

\newcommand{\bfp}{\boldsymbol{p}}
\newcommand{\bfq}{\boldsymbol{q}}

\newcommand{\bfx}{\boldsymbol{x}}

\newcommand{\bfC}{\boldsymbol{C}}
\newcommand{\bfD}{\boldsymbol{D}}

\newcommand{\bfM}{\boldsymbol{M}}

\newfont{\tenbfsl}{cmbxti9 scaled 1200}% <-- for idem tensor
\newcommand{\id}{\mbox{\tenbfsl 1\/}}

\newcommand{\bfvarphi}{\boldsymbol{\varphi}}

\newcommand{\bfxi}{\boldsymbol{\xi}}

\newcommand{\bfnu}{\boldsymbol{\nu}}
\newcommand{\bftau}{\boldsymbol{\tau}}

\newcommand{\da}{\,\text{d}a}
\newcommand{\dv}{\,\text{d}v}

\newcommand{\calA}{\mathcal{A}}
\newcommand{\calB}{\mathcal{B}}

\newcommand{\calD}{\mathcal{D}}
\newcommand{\calF}{\mathcal{F}}

\newcommand{\calL}{\mathcal{L}}

\newcommand{\calP}{\mathcal{P}}

\newcommand{\calS}{\mathcal{S}}
\newcommand{\calV}{\mathcal{V}}
\newcommand{\calW}{\mathcal{W}}

\newcommand{\bdy}{\calB}

\newcommand{\Grad}{\hbox{\rm grad}\mskip2mu}
\newcommand{\Div}{\hbox{\rm div}\mskip2mu}

\newcommand{\prt}{\calP}

\newcommand{\ep}{\varepsilon}

\newfont{\tenbbb}{msbm10}
\newfont{\svnbbb}{msbm8}

\newcommand{\bbN}{\mbox{\tenbbb N\/}}

\newcommand{\xis}{\xi_{\scriptscriptstyle\calS}}

\newtheoremstyle{slplain}% name
  {.5\baselineskip\@plus.2\baselineskip\@minus.2\baselineskip}% Space above
  {.5\baselineskip\@plus.2\baselineskip\@minus.2\baselineskip}% Space below
  {\slshape}% Body font
  {}%Indent amount (empty = no indent, \parindent = para indent)
  {\bfseries}%  Thm head font
  {.}%       Punctuation after thm head
  { }%      Space after thm head: " " = normal interword space;
  {}%       Thm head spec

\theoremstyle{slplain}
\newtheorem{rmk}{Remark}

\begin{document}

\title{Extended Larch{\'e}--Cahn framework for reactive Cahn--Hilliard multicomponent systems}
\author{Santiago P.\ Clavijo$^1$, Luis Espath$^2$ \& Victor M.\ Calo$^3$}

\address{$^1$Ali I. Al-Naimi Petroleum Engineering Research Center, King Abdullah University of Science \& Technology (KAUST), Thuwal, 23955-6900, Saudi Arabia.}
\email{$^1$sapenacl91@gmail.com }
\address{$^2$Department of Mathematics, RWTH Aachen University, Pontdriesch 14-16, 52062 Aachen, Germany.}
\email{$^2$espath@gmail.com}
\address{$^3$School of Earth and Planetary Sciences, Curtin University, Kent Street, Bentley, Perth, WA 6102, Australia.}
\address{\phantom{$^3$}Curtin Institute for Computation, Curtin University, Kent Street, Bentley, Perth, WA 6102, Australia.}
\address{\phantom{$^3$}Mineral Resources, Commonwealth Scientific and Industrial Research Organisation (CSIRO), 10 Kensington, Perth, WA 6152, Australia.}
\email{$^3$vmcalo@gmail.com}

\date{\today}
\maketitle

\begin{abstract}

At high temperature and pressure, solid diffusion and chemical reactions between rock minerals lead to phase transformations. Chemical transport during uphill diffusion causes phase separation, that is, spinodal decomposition. Thus, to describe the coarsening kinetics of the exsolution microstructure, we derive a thermodynamically consistent continuum theory for the multicomponent Cahn--Hilliard equations while accounting for multiple chemical reactions and neglecting deformations. Our approach considers multiple balances of microforces augmented by multiple constituent content balance equations within an extended Larch{\'e}--Cahn framework. As for the Larch{\'e}--Cahn framework, we incorporate into the theory the Larch{\'e}--Cahn derivatives with respect to the phase fields and their gradients. We also explain the implications of the resulting constrained gradients of the phase fields in the form of the gradient energy coefficients. Moreover, we derive a configurational balance that includes all the associated configurational fields in agreement with the Larch{\'e}--Cahn framework. We study phase separation in a three-component system whose microstructural evolution depends upon the reaction-diffusion interactions and to analyze the underlying configurational fields. This simulation portrays the interleaving between the reaction and diffusion processes and how the configurational tractions drive the motion of interfaces.

\textbf{AMS subject classifications:}
$\cdot$
74N20 % Mechanics of deformable solids - Dynamics of phase boundaries
$\cdot$
80A22 % Classical thermodynamics, heat transfer - Stefan problems, phase changes, etc.
$\cdot$
80A17 % Classical thermodynamics, heat transfer - Thermodynamics of continua
$\cdot$
82C26 % Statistical mechanics, structure of matter - Dynamic and nonequilibrium phase transitions (general)
$\cdot$
35L65 % Partial differential equations - Conservation laws
$\cdot$
\end{abstract}

\tableofcontents

\section{Introduction}

Deep in the Earth, both high temperature and pressure allow for solid diffusion and chemical reactions between rock minerals, which in turn, lead to phase transformations and induced deformation. Significantly, the transport of chemical constituents during uphill diffusion generates phase separation processes as a result of the geothermal gradient in the crust. The phases that compose these solid solutions of minerals diffuse at different rates, and when considering changes in temperature as a result of uphill diffusion, for instance during cooling, phase separation processes such as spinodal decomposition occur. For example, ternary feldspars formed by orthoclase, anorthite, and albite show spinodal decomposition during cooling. Thus, such a process controls the coarsening kinetics of the exsolution microstructure~\cite{yund1974coarsening,petrishcheva2012exsolution, petrishcheva2020spinodal}. Rocks are complex systems composed of several minerals, grain-boundaries, fractures, and pore space where the chemical and mechanical properties may vary in each direction. We describe each mineral as a component of a solid solution; this interpretation of the mixture allows us to explain the coupled reactive spinodal decomposition during exsolution. As a case study, we model a phase merging process driven by interfacial responses coupled with chemical reactions as a first attempt to understand the dynamics of reactive exsolution by spinodal decomposition~\cite{abart2009exsolution}. We derive a thermodynamically-consistent reactive $n$ species Cahn-Hilliard model that captures the dynamics of such interactions while following in detail the configurational forces that drive this coupled kinematical process.

The multicomponent Cahn--Hilliard model is a useful tool for studying the kinetics of multiphase systems undergoing phase separation. Most importantly, this model tracks the microstructure evolution of the resulting phases to enhance our understanding of the resulting material properties. To describe the underlying physics of this problem, we consider $n$ phase fields representing the concentration of conserved species and use a set of coupled Cahn--Hilliard equations. This representation leads to a system of $n$ degenerate nonlinear fourth-order parabolic partial differential equations. The degeneracy is due to a nonlinear mobility tensor that can vanish depending on the phase field values. We assume that there exist $n$ microforce balances, as similarly proposed by Fried \& Gurtin \cite{Fri93,Fri94}, and $n$ mass balances accounting for all the relevant chemical reactions, as similarly proposed by Clavijo et al. \cite{Cla19}. We then build an extended Larch{\'e}--Cahn framework to account for the interdependence between the conserved species. Given the set $\bfvarphi = \{ \varphi^1, \ldots, \varphi^n\}$ of species, where $n\in\bbN$, we consider $n-1$ independent species $\tilde{\bfvarphi} = \bfvarphi \setminus \{ \varphi^\sigma \}$ while the $\sigma$-th conserved species is used as a reference and determined by $\tilde{\bfvarphi}$. Thus, to compute partial derivatives with respect to $\varphi^\alpha$ and $\Grad\varphi^\alpha$, the dependence among species must be taken into account. We then redefine the partial derivative of functions depending upon $\bfvarphi$ and $\Grad\bfvarphi$ where $\varphi^\sigma$ is constrained to derive the multicomponent Cahn--Hilliard equations. Moreover, in defining these partial derivatives, we arrive at a \emph{constrained} inner product on a constrained space to appropriately define the gradient energy coefficients $\Gamma^{\alpha\beta}$.

The outline of this article is as follows. In section~\S\ref{sc:framework}, we introduce the balances of microforces and augment them with mass balances. In section~\S\ref{sc:configurational.forces}, we present the configurational forces and their balances, and describe how they drive the interface evolution. In section~\S\ref{sc:dimensionless}, we make the equations dimensionless. Section~\S\ref{sc:simulations} exemplifies the use of configurational tractions to explain the evolution of a three-alloy mixture. The final section enumerates our conclusions and future work. Appendix \ref{ap:all} presents the mathematical derivations of this theory.

\section{Theoretical framework}\label{sc:framework}

We give a brief overview of the theoretical framework that describes the isothermal evolution of $n\ge2$ reacting and diffusing chemical constituents that occupy a fixed region $\bdy$ of a three-dimensional point space.

\subsection{Constituent content balances}

We assume that a mass density $\varrho^\alpha$, a diffusive flux $\bfj^\alpha$, and a reactive mass supply rate $s^\alpha$ characterize the instantaneous state of each constituent $\alpha=1,\dots,n$. Also, we require that $\varrho^\alpha$, $\bfj^\alpha$, and $s^\alpha$ evolve subject to a pointwise constituent content balance in the form
\begin{equation}\label{eq:massbal.pointwise.precursor}
\dot{\varrho}^\alpha=-\Div\bfj^\alpha+s^\alpha,\qquad\forall\,1\le\alpha\le{n},
\end{equation}
where a superposed dot denotes partial differentiation with respect to time and $\text{div}$ denotes the divergence on $\bdy$. Stipulating that the mass supply rates and the diffusive fluxes satisfy constraints of the form
\begin{equation}\label{eq:constraint.mass.fluxes.1}
\sum _{\alpha=1}^{n} s^\alpha = 0
\qquad \text{and} \qquad
\sum _{\alpha=1}^{n} \bfj^\alpha = 0,
\end{equation}
we sum the constituent content balance~\eqref{eq:massbal.pointwise.precursor} over $\alpha$ from $1$ to $n$ to find that the total mass density
\begin{equation}\label{eq:totaldensity.def}
\varrho=\sum_{\alpha=1}^n\varrho^\alpha
\end{equation}
must satisfy $\dot{\varrho}=0$ and, thus, is constant.

\subsection{Concentrations, phase fields, and microforce balances}\label{sc:fundamental}

Introducing a concentration
\begin{equation}\label{eq:concetration.def}
\varphi^\alpha=\frac{\varrho^\alpha}{\varrho}
\end{equation}
for each species $\alpha=1,\dots,n$, from expressions~\eqref{eq:totaldensity.def} and \eqref{eq:concetration.def} together with the requirement that the total mass density $\varrho$ is a fixed constant, we have that the following constraint
\begin{equation}\label{eq:const}
\sum_{\alpha=1}^n\varphi^\alpha = 1
\end{equation}
must hold in conjunction with~\eqref{eq:constraint.mass.fluxes.1}. Moreover, from constituent content balance~\eqref{eq:massbal.pointwise.precursor} for constituent $\alpha=1,\dots,n$, we have that
\begin{equation}\label{eq:massbal.pointwise.nosupply}
\varrho\mskip1mu\dot{\varphi}^\alpha=-\Div\bfj^\alpha+s^\alpha,\qquad\forall\,1\le\alpha\le{n}.
\end{equation}

Next, let $\bfxi^\alpha$ be the $\alpha$-th microstress, and $\pi^\alpha$ ($\gamma^\alpha$) field is the $\alpha$-th internal (external) microforce. Thus, we express the microforce balances of Fried \& Gurtin \cite[\S{IV}]{fried1999coherent} in its pointwise form as
\begin{equation}\label{eq:balmicro.pointwise}
\Div\bfxi^\alpha + \pi^\alpha + \gamma^\alpha = 0.
\end{equation}
In the partwise form of expression~\eqref{eq:balmicro.pointwise}, the surface microtraction is $\xi^\alpha\coloneqq\bfxi^\alpha\cdot\bfn$.

\subsection{Thermocompatible constitutive relations}

We introduce constitutive relations for the diffusive flux $\bfj^\alpha$, the reactive mass supply rate $s^\alpha$, the internal microforce density $\pi^\alpha$, and the microstress $\bfxi^\alpha$ for each constituent $\alpha=1,\dots,n$, which allow us to close the system of evolution equations for the phase fields $\varphi^\alpha$, $\alpha=1,\dots,n$.  These relations must be compatible with the constraints~\eqref{eq:constraint.mass.fluxes.1} and~\eqref{eq:const} and with the first and second laws of thermodynamics, which, since we consider only isothermal processes, combine to yield an inequality of the form
\begin{equation}\label{eq:pointwise.free.energy.imbalance.isothermal}
\varrho\dot\psi-\sum_{\alpha=1}^{n}\{(\varrho\mskip1mu\mu^{\alpha}-\pi^\alpha)\dot\varphi^\alpha-\bfj^\alpha\cdot\Grad\mu^{\alpha}+\mu^\alpha s^\alpha
+\bfxi^\alpha\cdot\Grad\dot\varphi^\alpha\}\le0,
\end{equation}
where $\psi$ is the specific free-energy, and $\mu^\alpha$ is the chemical potential of constituent $\alpha$ (for details, see Appendix~\ref{sc:thermodynamics}). We define the chemical potential using the Coleman--Noll procedure in the next section.

\subsection{Thermodynamical constraints}

Throughout the derivation of the constitutive relations for the multicomponent Cahn--Hilliard system, we use the Larch{\'e}--Cahn derivative~\eqref{eq:lcder} from Appendix~\ref{sc:larchecahnderivaties}. Using the Coleman--Noll procedure~\cite{ Col63}, we find the sufficient conditions to ensure the inequality~\eqref{eq:pointwise.free.energy.imbalance.isothermal} for arbitrary fields. Thus, a set of paired constitutive equations emerges for each kinematic process. We assume the following constitutive dependency of the free energy $\psi$ within the context of isothermal processes
\begin{equation}\label{eq:constfree}
\psi \coloneqq \hat{\psi}(\bfvarphi, \Grad\bfvarphi),
\end{equation}
which specializes the free-energy~\eqref{eq:pointwise.free.energy.imbalance.isothermal} as follows
\begin{equation}\label{eq:free.coleman}
\sum_{\alpha=1}^{n} \left\{\varrho\mskip2mu\mu^{\alpha} - \pi^\alpha - \varrho\mskip2mu\frac{\partial^{(\sigma)} \hat{\psi}}{{\partial \varphi^\alpha}} \right\} \dot{\varphi}^\alpha + \sum_{\alpha=1}^{n} \left\{\bfxi^\alpha - \varrho\mskip2mu\frac{\partial^{(\sigma)} \hat{\psi}}{\partial (\Grad \varphi^\alpha)} \right\} \cdot \Grad \dot{\varphi}^\alpha - \sum_{\alpha=1}^{n} \left\{ \bfj^\alpha \cdot \Grad \mu^{\alpha} + \mu^\alpha s^\alpha \right\} \leq0.
\end{equation}
The free-energy imbalance~\eqref{eq:free.coleman} must hold for any arbitrary $\dot{\varphi}^\alpha$, $\Grad\dot{\varphi}^\alpha$, and $\Grad\mu^\alpha$ fields at a given time and place. Thus, the following relations must hold
\begin{subequations}\label{eq:consitutive}
\begin{align}
\pi^\alpha_\sigma & = \varrho\mskip2mu\left(\mu^\alpha_\sigma  - \frac{\partial^{(\sigma)} \hat{\psi}}{{\partial \varphi^\alpha}}\right), \label{eq:microforcebal} \\
\bfxi^\alpha_\sigma & = \varrho\mskip2mu\frac{\partial^{(\sigma)} \hat{\psi}}{\partial (\Grad \varphi^\alpha)}, \label{eq:microstressbal} \\[6pt]
\bfj^\alpha_\sigma & = -\sum_{\beta=1}^{n}\bfM^{\alpha\beta} \, \Grad \mu^\beta_\sigma, \label{eq:massflux}
\end{align}
\end{subequations}
where $\bfM$ is the mobility tensor, which must be positive semi-definite, that is, $\sum_{\alpha=1}^n\sum_{\beta=1}^{n}\bfp^\alpha\cdot\bfM^{\alpha\beta}\bfp^\beta\ge0$, holds for all $\bfp$. As~\eqref{eq:free.coleman} expresses all thermodynamically consistent choices, we write the terms $\pi^\alpha\coloneqq\pi^\alpha_\sigma$, $\mu^\alpha\coloneqq\mu^\alpha_\sigma$, and $\bfxi^\alpha\coloneqq\bfxi^\alpha_\sigma$ relative to the Larch{\'e}--Cahn construction given their explicit dependence on the Larch{\'e}--Cahn derivatives as an essential consequence of~\eqref{eq:const}. As a byproduct, we also write the mass flux, $\bfj^\alpha\coloneqq\bfj^\alpha_\sigma(\bfx,t;\Grad\mu^\alpha_\sigma)$, and the surface microtraction, $\xis^\alpha\coloneqq{\xis^\alpha}_{\!\sigma}(\bfx,t;\bfxi^\alpha_\sigma)$ as constructions dependent on the Larch{\'e}--Cahn derivative. Finally, intrinsically in these definitions, we express all quantities relative to the $\sigma$-th species.

Guided by the original Cahn--Hilliard equation~\cite{ cahn1958free}, we assume that the evolution of the Ginzburg--Landau free energy governs the nature of phase separation undergoing spinodal decomposition. In a multicomponent system, we determine the constitutive relations in~\eqref{eq:consitutive} from the Ginzburg-Landau free energy expressed as
\begin{equation}\label{eq:constfree1}
\hat{\psi}(\bfvarphi, \Grad \bfvarphi) = N_v k_B \vartheta \left( \sum _{\alpha=1} ^{n} \, \varphi^\alpha \ln \varphi^\alpha \right) + N_v \sum _{\alpha=1}^{n} \sum _{\beta=1} ^{n} \Omega^{\alpha\beta} \varphi^\alpha \varphi^\beta + \dfrac{1}{2} \sum _{\alpha=1} ^{n} \sum _{\beta=1} ^{n} \Gamma^{\alpha\beta} \, \Grad \varphi^\alpha \cdot \Grad \varphi^{\beta},
\end{equation}
where $N_v$ is the total number of molecules of the species $\alpha$ per unit volume, $k_B$ is the Boltzmann constant, and $\Omega^{\alpha\beta}$ represents the interaction energy between the mass fraction of the $\alpha$-th and $\beta$-th species, which is reciprocal; thus, $\Omega^{\alpha\beta}$ is symmetric. The interaction energy is positive and is related to the critical temperature for each pair of species, $\vartheta^{\alpha\beta}_c$, (between the $\alpha$-th and $\beta$-th species). Following standard convention, we adopt that $\Omega^{\alpha\beta} = 0$ when $\alpha = \beta$ and $\Omega^{\alpha\beta} = 2 k_{B}\vartheta^{\alpha\beta}_c$ when $\alpha \neq \beta$~\cite{elliott1997diffusional,gurtin1989nonequilibrium,cahn1958free}. Furthermore, $\Gamma^{\alpha\beta} = \sigma^{\alpha\beta}\ell^{\alpha\beta}$ [force] (no sum on $\alpha$ and $\beta$) represents the magnitude of the interfacial energy between the $\alpha$-th and $\beta$-th species. The parameters $\sigma^{\alpha\beta}$ and $ \ell^{\alpha\beta}$ are the interfacial tension [force/length] and the interfacial thickness\footnote{This expression corresponds to the root mean square effective "interaction distance", as suggested by Cahn \& Hilliard \cite{cahn1958free}.} for each pair of species (between the $\alpha$-th and $\beta$-th species) [length], respectively. Cahn \& Hilliard \cite{cahn1958free} define the force $\Gamma^{\alpha\beta}$ as $N_v\Omega^{\alpha\beta} (\ell^{\alpha\beta})^{2}$.

We express the relative chemical potential of the $\alpha$-th species, in the Larch{\'e}--Cahn sense, by combining the expressions~\eqref{eq:microforcebal},~\eqref{eq:microstressbal}, the microforce balance~\eqref{eq:balmicro.pointwise}, and the constitutive relation for the free energy~\eqref{eq:constfree1}, we arrive at
\begin{equation}\label{eq:chemicalpot1}
\mu^\alpha_\sigma = \frac{\partial^{(\sigma)} \hat{\psi}}{\partial \varphi^\alpha} - \Div \frac{\partial^{(\sigma)} \hat{\psi}}{\partial (\Grad \varphi^\alpha)} - \frac{1}{\varrho}(\gamma^{\alpha} - \gamma^{\sigma}).
\end{equation}
Therefore, the combination of~\eqref{eq:chemicalpot1} with~\eqref{eq:constfree1} specializes to
\begin{equation}\label{eq:chemicalpot2}
\mu^\alpha_\sigma = N_v k_{B} \vartheta\ \ln \left( \dfrac{\varphi^\alpha}{\varphi^\sigma} \right) + 2 N_v \sum_{\beta=1}^{n} (\Omega^{\alpha\beta} - \Omega^{\sigma\beta}) \varphi^\beta - \sum _{\beta=1} ^{n} (\Gamma^{\alpha\beta} - \Gamma^{\sigma\beta}) \, \Div \Grad \varphi^\beta - \frac{1}{\varrho}(\gamma^{\alpha} - \gamma^{\sigma}).
\end{equation}

In the following, we assume an isotropic mobility $\bfM^{\alpha\beta}=M^{\alpha\beta}\id$ being $\bfM^{\alpha\beta}$ symmetric, but we consider the off-diagonal terms in the Onsager reciprocal relations. We use the standard assumption that the mobility coefficients depend on the phase composition. In particular, we express this dependency in terms of the concentration of each species. We use the definition $M^{\alpha\beta} \coloneqq M_0^{\alpha\beta} \varphi^\alpha(\delta^{\alpha\beta} - \varphi^{\beta})$ with no summation on $\alpha$ and $\beta$ and $M_0^{\alpha\beta}$ is the mobility between the $\alpha$ and $\beta$ species, with dimension of length$^4$ per unit force and time~\cite{ elliott1997diffusional}. Thus,~\eqref{eq:constraint.mass.fluxes.1} implies the following relation
\begin{equation}\label{eq:constraint.mass.fluxes.2}
\sum _{\beta=1}^{n} M^{\alpha\beta} = 0, \quad \forall \alpha.
\end{equation}

\subsection{Chemical reaction}\label{sc:constitutive.equations}

Let $\varphi^\alpha$ be the concentration of a species $\calA^\alpha$, such that $\varphi^\alpha \coloneqq [\calA^\alpha]$. Following Krambeck~\cite{ krambeck1970mathematical}, we express the $c$-th chemical reaction, in a set of $n_s$ chemical reactions, $n_s\in\bbN$, as
\begin{equation}\label{eq:several.chemical.reaction}
\sum _{\alpha=1} ^n \upsilon_r^{c,\alpha} \calA^\alpha \, \mathop{\rightleftharpoons} _{k_-^c} ^{k_+^c} \, \sum _{\alpha=1} ^n \upsilon_p^{c,\alpha} \calA^\alpha,\qquad\forall\,1\le{c}\le{n_s},
\end{equation}
where $\upsilon_r^{c,\alpha}$ and $\upsilon_p^{c,\alpha}$ are the $\alpha$-th stoichiometric coefficient of the $c$-th chemical reaction of the reactants and products, respectively. The number of non-zero stoichiometric coefficients $\upsilon_r^{c,\alpha}$ and $\upsilon_p^{c,\alpha}$ define the number of reactants $n_r^c$ and products $n_p^c$ in the $c$-th chemical reaction. For the $c$-th chemical reaction, zeros populate $\upsilon_r^{c,\alpha}$ for $\alpha > n_r^c$, whereas zeros populate $\upsilon_p^{c,\alpha}$ for $\alpha \le n_r^c$. $k_+^c$ ($k_-^c$)  denotes the $c$-th forward (backward) reaction rate (see, for details Appendix~\ref{sc:theoryofreactingmaterial}). We focus on ideal materials, then, the $c$-th rates of both the forward and backward reactions read
\begin{align}\label{eq:forward.reaction}
r_+^c &\coloneqq k_+^c \prod _{\alpha=1} ^n (\varphi^\alpha) ^{\upsilon_r^{c,\alpha}},\\
\label{eq:backward.reaction}
r_-^c &\coloneqq k_-^c \prod _{a=1} ^n (\varphi^\alpha) ^{\upsilon_p^{c,\alpha}}.
\end{align}
Finally, the internal rate of mass supply term for all $n_s$ chemical reactions that enters in~\eqref{eq:massbal.pointwise.nosupply} is
\begin{equation}\label{eq:forward.backward.reaction}
s^\alpha \coloneqq -\sum_{c=1}^{n_s}(\upsilon_r^{c,\alpha}-\upsilon_p^{c,\alpha})(r_+^c-r_-^c).
\end{equation}

\section{Configurational fields}\label{sc:configurational.forces}

We describe the interfacial evolution, and its thermodynamics using the configurational forces proposed by Gurtin \cite{Gur08}, which relate the integrity of the material and the movement of its defects. The configurational forces expend the power associated with the transfer of matter, which allow us to interpret them thermodynamically. Using the configurational balance for a part $\prt$ by Fried \cite{Fri06b}, we have
\begin{equation}
\int\limits_{\calS}\bfC\bfn\da+\int\limits_{\prt}(\bff+\bfe)\dv=\bf0,
\end{equation}
which renders, after localization,
\begin{equation}\label{eq:pointwise.configurational.balance}
\Div\bfC+\bff+\bfe=\bf0,
\end{equation}
where $\bfC$ is the configurational stress tensor and $\bff$ ($\bfe$) is the internal (external) force.

Following Appendix~\ref{sc:configurationalstresses}, we substitute the constitutive relation~\eqref{eq:zeta} in the relation~\eqref{eq:configurational.relation}, allows us to express the configurational stress as
\begin{equation}\label{eq:configurational.stress}
\bfC\coloneqq\varrho\left(\psi-\sum_{\alpha=1}^{n}\mu^\alpha\varphi^\alpha\right)\id-\sum_{\alpha=1}^{n}\Grad\varphi^\alpha\otimes\bfxi^\alpha.
\end{equation}
We obtain explicit forms for the internal and external configurational forces by combining~\eqref{eq:consitutive} and~\eqref{eq:pointwise.configurational.balance} with~\eqref{eq:configurational.stress}, that is
\begin{equation}\label{eq:configurational.forces}
\bff\coloneqq\sum_{\alpha=1}^{n}\varrho\mskip2mu\varphi^\alpha\mskip+2.5mu\Grad\mu^\alpha \qquad \text{and} \qquad \bfe\coloneqq-\sum_{\alpha=1}^{n}\gamma^\alpha\mskip+2.5mu\Grad\varphi^\alpha.
\end{equation}

By considering the Larch\'e-Cahn derivatives, we express the configurational stress~\eqref{eq:configurational.stress} as a configurational stress relative to the $\sigma$-th species as follows
\begin{equation}\label{eq:relative.configurational.stress}
\bfC_\sigma\coloneqq\varrho\left(\psi-\sum_{\alpha=1}^{n}\mu^\alpha_\sigma\varphi^\alpha\right)\id-\sum_{\alpha=1}^{n}\Grad\varphi^\alpha\otimes\bfxi^\alpha_\sigma,
\end{equation}
while
\begin{equation}\label{eq:relative.configurational.forces}
\bff_\sigma\coloneqq\sum_{\alpha=1}^{n}\varrho\mskip2mu\varphi^\alpha\mskip+2.5mu\Grad\mu^\alpha_\sigma,
\end{equation}
is the relative internal configurational force. The external configurational force is not determined using a constitutive relation; thus, it does not depend upon the choice of the reference species.

\begin{rmk}[\bf Invariance of configurational balance to reference species]
Let $\varphi^\sigma$ be the reference species. We establish the following relations for the terms appearing in the configurational stress~\eqref{eq:relative.configurational.stress}
\begin{align}\label{eq:id.1}
- \sum_{\alpha=1}^{n}\mu^\alpha_\sigma\varphi^\alpha &= - \sum_{\alpha=1}^{n}\mu^\alpha\varphi^\alpha + \mu^\sigma\sum_{\alpha=1}^{n}\varphi^\alpha %\nonumber \\
                                                     %&
                                                       = - \left(\sum_{\alpha=1}^{n}\mu^\alpha\varphi^\alpha\right) + \mu^\sigma,
\end{align}
and
\begin{align}\label{eq:id.2}
\sum_{\alpha=1}^{n}\Grad\varphi^\alpha\otimes\bfxi^\alpha_\sigma &= \sum_{\alpha=1}^{n}\Grad\varphi^\alpha\otimes(\bfxi^\alpha - \bfxi^\sigma) %\nonumber \\ &
= \sum_{\alpha=1}^{n}\Grad\varphi^\alpha\otimes\bfxi^\alpha - \left(\sum_{\alpha=1}^{n}\Grad\varphi^\alpha\right)\otimes\bfxi^\sigma \nonumber \\
&= \sum_{\alpha=1}^{n}\Grad\varphi^\alpha\otimes\bfxi^\alpha.
\end{align}
while for the internal configurational force~\eqref{eq:relative.configurational.forces}
\begin{align}\label{eq:id.3}
\sum_{\alpha=1}^{n}\varphi^\alpha\mskip+2.5mu\Grad\mu^\alpha_\sigma &= \sum_{\alpha=1}^{n}\varphi^\alpha\mskip+2.5mu\Grad\mu^\alpha - \Grad\mu^\sigma\sum_{\alpha=1}^{n}\varphi^\alpha%, \nonumber \\ &
                                                                      = \left(\sum_{\alpha=1}^{n}\varphi^\alpha\mskip+2.5mu\Grad\mu^\alpha\right) - \Grad\mu^\sigma.
\end{align}
Analyzing~\eqref{eq:id.1} and~\eqref{eq:id.2}, we conclude that only one term in $\bfC_\sigma$~\eqref{eq:relative.configurational.stress} depends on the reference species. Therefore, the relative configurational stress becomes
\begin{equation}\label{eq:relative.configurational.stress.rewritten}
\bfC_\sigma \coloneqq \bfC + \varrho\mskip2mu\mu^\sigma\id.
\end{equation}
Meanwhile, we specialize representation of the relative internal configurational force~\eqref{eq:relative.configurational.forces}  with~\eqref{eq:id.3} yielding
\begin{equation}\label{eq:relative.configurational.force.rewritten}
\bff_\sigma \coloneqq \bff - \varrho\mskip2mu\Grad\mu^\sigma,
\end{equation}
Finally, although both the configurational stress and the internal configurational force explicitly depend on the choice $\varphi^\sigma$; nevertheless, their dependencies cancel each other's contribution to the configurational balance~\eqref{eq:pointwise.configurational.balance},
\begin{align}\label{eq:relative.pointwise.configurational.balance}
\Div\bfC_\sigma + \bff_\sigma &= \Div(\bfC + \varrho\mskip2mu\mu^\sigma\id) + \bff - \varrho\mskip2mu\Grad\mu^\sigma, \nonumber \\
&= \Div\bfC + \bff.
\end{align}
\qed
\end{rmk}

\section{Dimensionless multicomponent Cahn--Hilliard equations}\label{sc:dimensionless}

The final system resulting from~\eqref{eq:massbal.pointwise.nosupply},~\eqref{eq:consitutive},~\eqref{eq:chemicalpot2}, and~\eqref{eq:forward.reaction}-\eqref{eq:forward.backward.reaction} reads
\begin{spreadlines}{0.75em}
\begin{equation}\label{eq:gov.eq}
\left\{\,
\begin{aligned}
\dot{\varphi}^\alpha = & s^\alpha - \Div\bfj^\alpha_\sigma,\\
\bfj^\alpha_\sigma = &-\sum_{\beta=1}^{n} M_0^{\alpha\beta} \varphi^\alpha(\delta^{\alpha\beta} - \varphi^{\beta}) \, \Grad \mu^\beta_\sigma,\\
\mu^\beta_\sigma = & N_v k_{B} \vartheta \ln \dfrac{\varphi^\beta}{\varphi^\sigma} + 2 N_v \sum_{\alpha=1}^{n} (\Omega^{\beta\alpha} - \Omega^{\sigma\alpha}) \varphi^\alpha - \sum _{\alpha=1} ^{n} (\Gamma^{\beta\alpha} - \Gamma^{\sigma\alpha}) \, \Div \Grad \varphi^\alpha - (\gamma^\beta + \gamma^\sigma),\\
s^\alpha = &-\sum_{c=1}^{n_s}\left\{(\upsilon^{c,\alpha}-\varpi^{c,\alpha})(k^c_+ \prod _{a=1} ^{n} (\varphi^a) ^{\upsilon ^{ca}}-k^c_- \prod _{a=1} ^{n} (\varphi^a) ^{\varpi ^{ca}})\right\},
\end{aligned}
\right.
\end{equation}
\end{spreadlines}
 $\text{in}\,\,\calD\times(0,T)$ with
\begin{spreadlines}{0.75em}
\begin{equation}\label{eq:gov.eq.bvp}
\left\{\,
\begin{aligned}
&\varphi^\alpha(\bfx,0)=\varphi^\alpha_0,&\text{in}\,\calD,\\
&\text{subject to periodic boundary conditions}&\text{on}\,\,\partial\calD\times(0,T),
\end{aligned}
\right.
\end{equation}
\end{spreadlines}
 where $\calD$ is the domain of interest.

To make the equations dimensionless, we introduce the reference energy density $\psi_0 \coloneqq 2 N_v k_{B} \vartheta$ and define the set of diffusion coefficients $\bfD^{\alpha\beta}$,
\begin{equation}
\bfD^{\alpha\beta} = \psi_0 M_0^{\alpha\beta} \varphi^{\alpha}(\delta^{\alpha\beta} - \varphi^{\beta})\qquad\text{no sum on $\alpha$ and $\beta$}.
\end{equation}
The reference energy density relates the species mobilities with the species diffusion as proposed in \cite{blickle2007einstein,einstein1905molekularkinetischen}. We also define the following dimensionless variables
\begin{equation}\label{eq:dimvar}
\overline{\bfx} = L_0^{-1}\bfx, \qquad \overline{t} = T_0^{-1}t, \qquad \overline{\vartheta}_c^{\alpha\beta} = \vartheta^{-1} \vartheta_c^{\alpha\beta}.
\end{equation}
Conventionally, the definition of the reference time $T_0$ for the Cahn--Hilliard system relates the diffusion coefficient, the interfacial thickness, and domain length, that is, $T_0 = D_0 \ell_0^2 L_0^{-4}$ where $L_0 \gg \ell_0$~\cite{ langtangen2016scaling, gomez2008isogeometric}. We set $D_0$ and $\ell_0$ as the reference diffusion coefficient and interface thickness of a reference species, and introduce the following dimensionless numbers for the multicomponent system
\begin{equation}\label{eq:dimlessnum}
\left\{\,
\begin{aligned}
\bar{k}^{+}_{c} &= k^{+}_c D_0^{-1} \ell^{-2}_0 L_0^4, \qquad \bar{k}^{-}_{c} = k^{-}_c D_0^{-1} \ell^{-2}_0 L_0^4, \qquad \overline{\psi} = \psi_0^{-1}\psi, \\[4pt]
\bar{\sigma}^{\alpha\beta} &= \sigma^{\alpha\beta}(\psi_0 L_0 )^{-1}, \qquad \bar{\bfD}^{\alpha\beta} =  \bfD^{\alpha\beta} D_0^{-1} \ell_0^{-2} L_0^{2}, \qquad \bar{\ell}^{\alpha\beta} = L_0^{-1}\ell^{\alpha\beta}, \quad \bar{\gamma}^\alpha = \psi_0^{-1}\gamma^\alpha.
\end{aligned}
\right.
\end{equation}
Thus, by inserting the dimensionless quantities in~\eqref{eq:gov.eq}, we find the following dimensionless forms
\begin{spreadlines}{0.75em}
\begin{equation}\label{eq:gov.eq.dimless}
\left\{\,
\begin{aligned}
\dot{\varphi}^\alpha = & \overline{s}^\alpha - \overline{\Div}\bar{\bfj}^\alpha_\sigma,\\
\bar{\bfj}^\alpha_\sigma = &-\sum_{\beta=1}^{n} \bar{\bfD}^{\alpha\beta} \, \overline{\Grad} \bar{\mu}^\beta_\sigma,\\
\bar{\mu}^\beta_\sigma = & \frac{1}{2} \ln \dfrac{\varphi^\beta}{\varphi^\sigma} + 2 \sum_{\alpha=1}^{n} (\bar{\vartheta}_c^{\beta\alpha} - \bar{\vartheta}_c^{\sigma\alpha}) \varphi^\alpha - \sum _{\alpha=1} ^{n} (\bar{\sigma}^{\beta\alpha}\bar{\ell}^{\beta\alpha} - \bar{\sigma}^{\sigma\alpha}\bar{\ell}^{\sigma\alpha}) \, \overline{\Div} \overline{\Grad} \varphi^\alpha - (\bar{\gamma}^\beta - \bar{\gamma}^\sigma),\\
\bar{s}^\alpha_{\text{int}} = &-\sum_{c=1}^{n_s}\left\{(\upsilon^{c,\alpha}-\varpi^{c,\alpha})(\bar{k}^c_+ \prod _{a=1} ^{n} (\varphi^a) ^{\upsilon ^{ca}}-\bar{k}^c_- \prod _{a=1} ^{n} (\varphi^a) ^{\varpi ^{ca}})\right\},
\end{aligned}
\right.
\end{equation}
\end{spreadlines}
$\text{in}\,\,\calD\times(0,T),$  with the initial condition~\eqref{eq:gov.eq.bvp}.

\section{Numerical simulation: merging of circular inclusions}\label{sc:simulations}

\begin{figure}[!htb]
  \centering
  \subfloat{\includegraphics[width=0.25\textwidth,trim={0px 0px 0px 0px},clip]{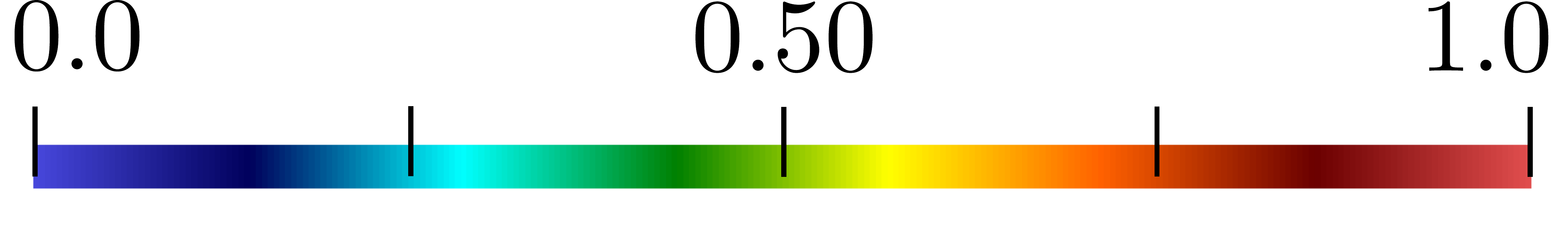}} \hspace{0.275cm}
  \subfloat{\includegraphics[width=0.25\textwidth,trim={0px 0px 0px 0px},clip]{pallete_num_12}} \hspace{0.275cm}
  \subfloat{\includegraphics[width=0.25\textwidth,trim={0px 0px 0px 0px},clip]{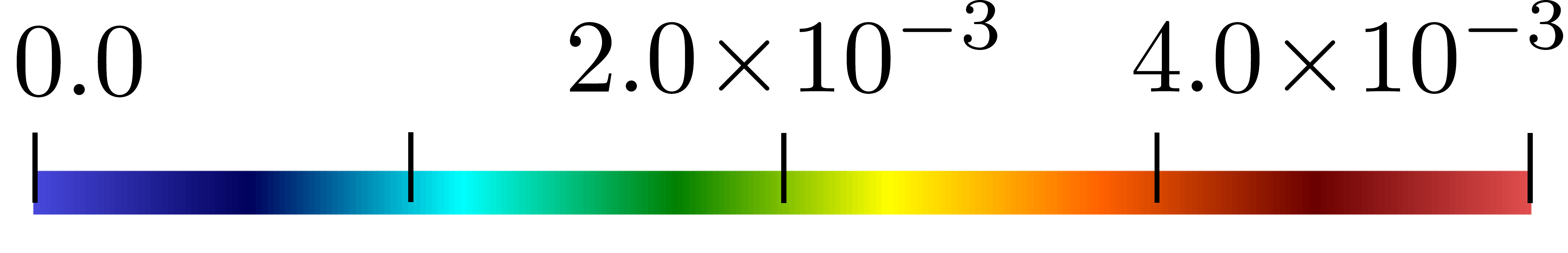}} \vspace{-0.4cm}\\
  \subfloat{\label{fg:01a}\includegraphics[height=0.275\textwidth,trim={0px 0px 0px 0px},clip]{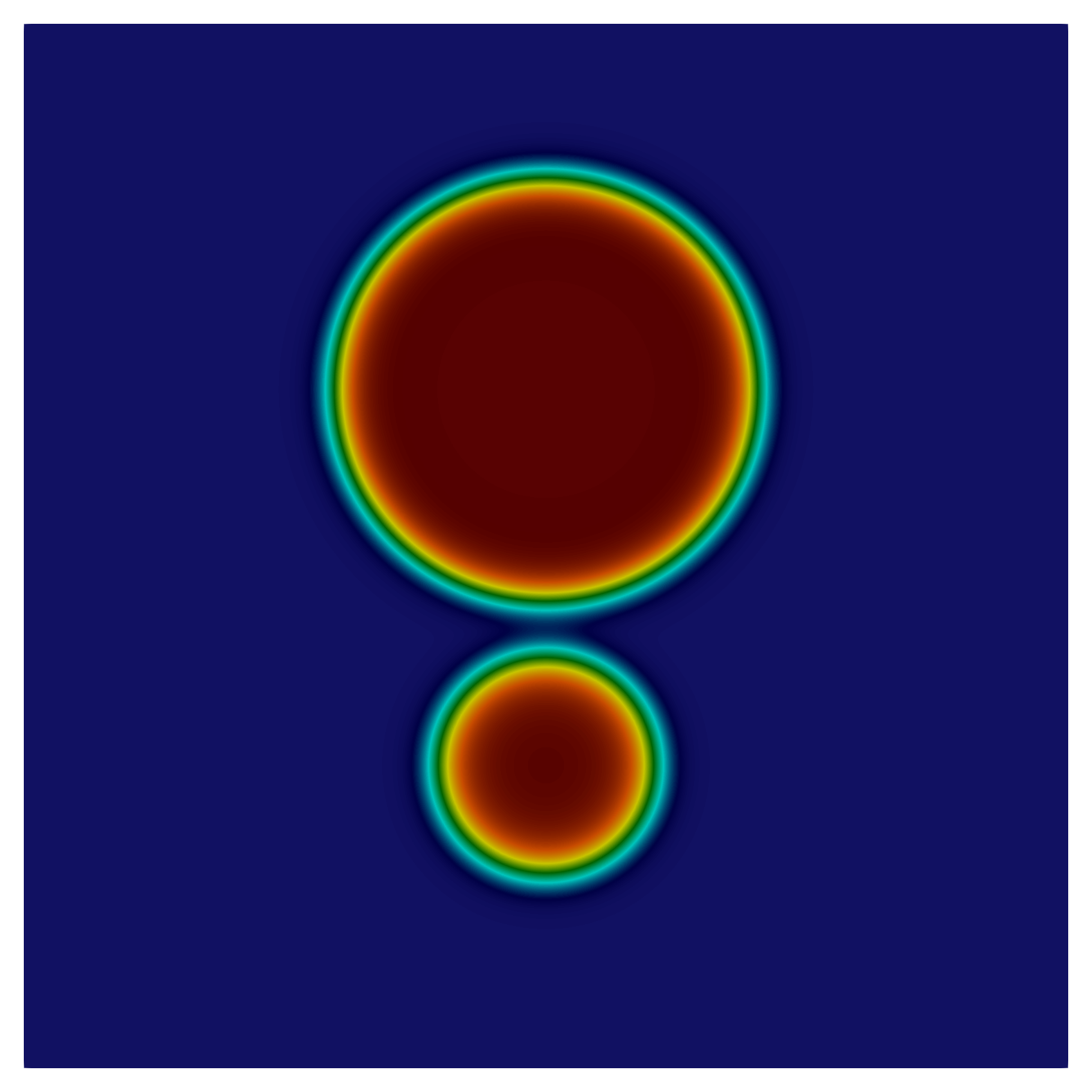}}
  \subfloat{\label{fg:01b}\includegraphics[height=0.275\textwidth,trim={0px 0px 0px 0px},clip]{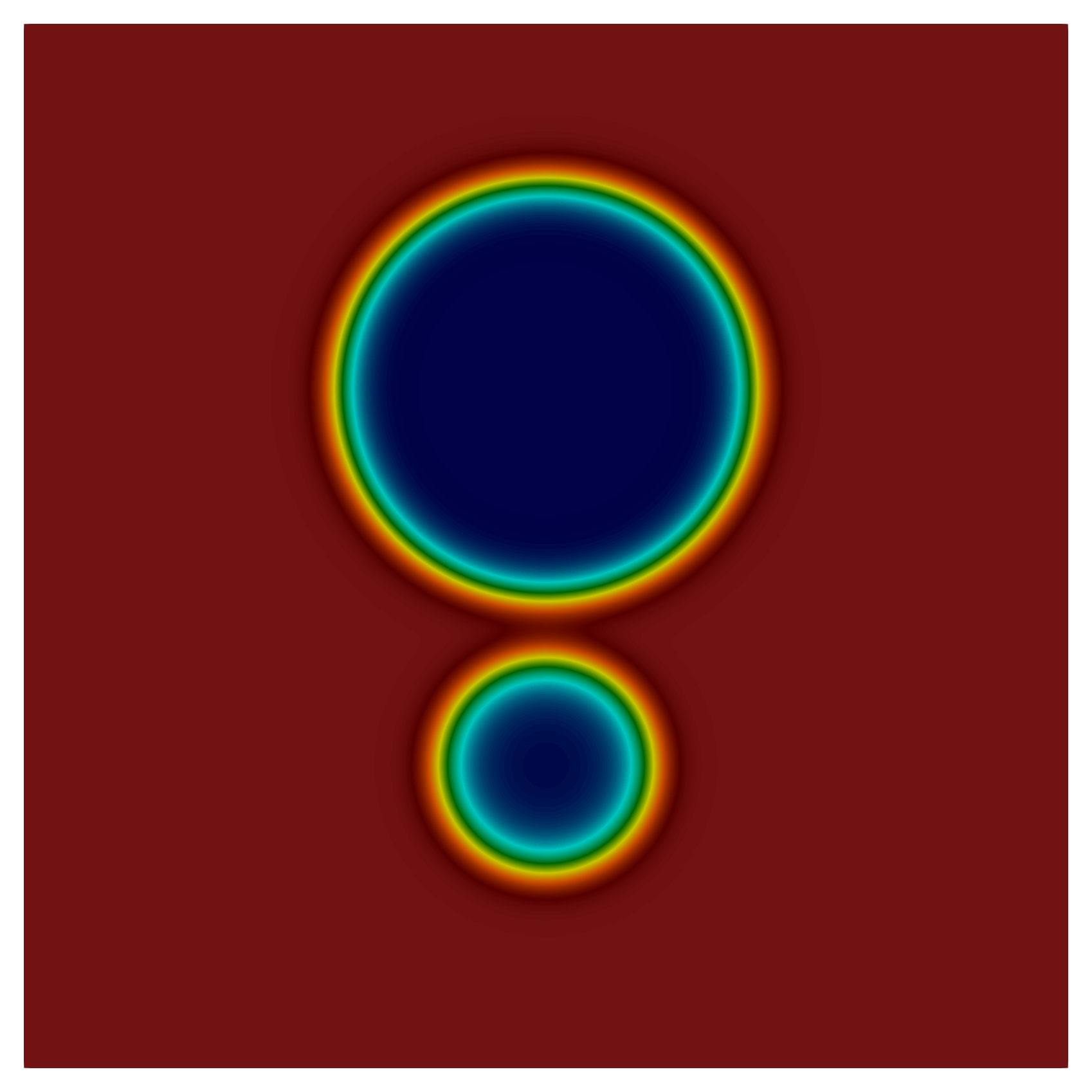}}
  \subfloat{\label{fg:01c}\includegraphics[height=0.275\textwidth,trim={0px 0px 0px 0px},clip]{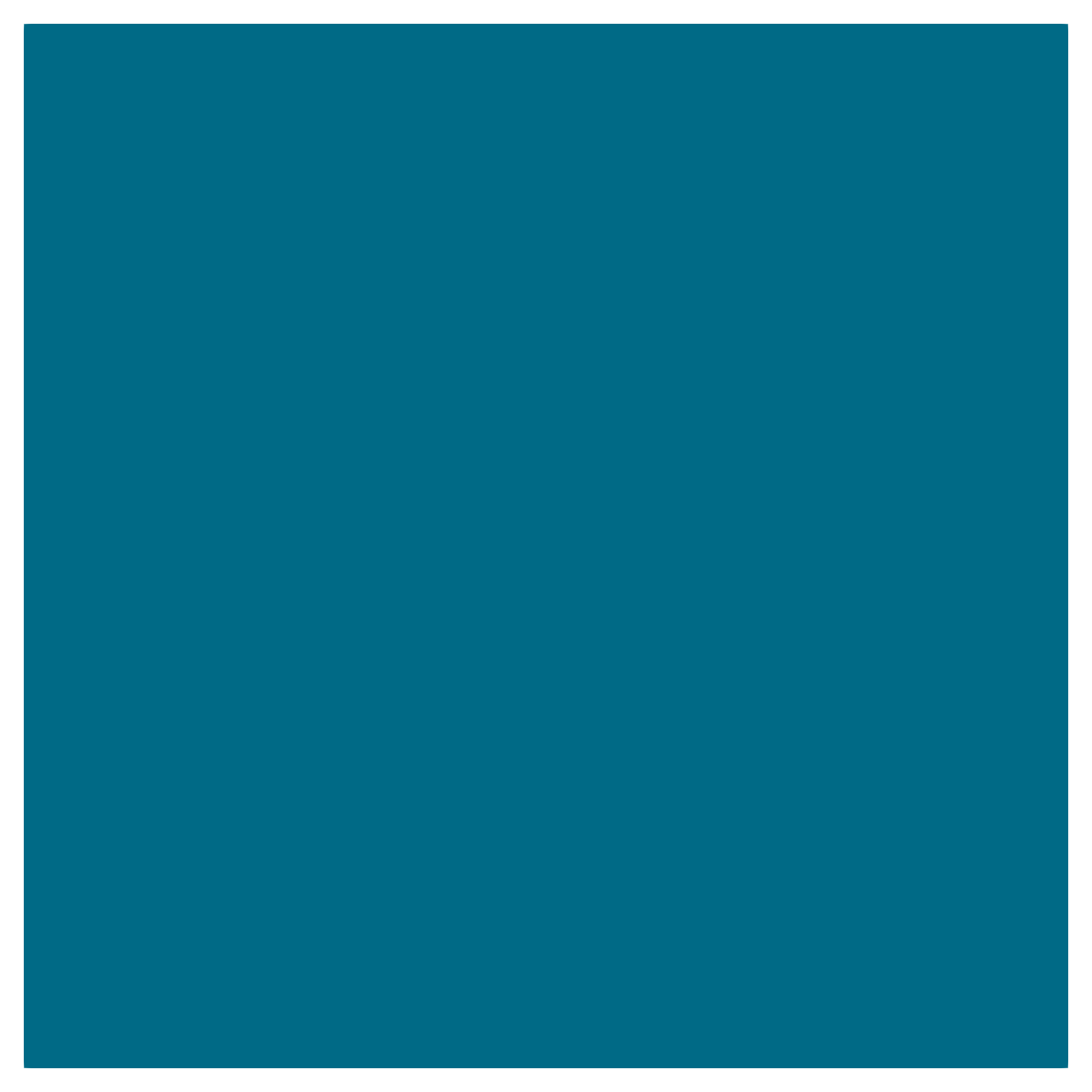}} \vspace{-0.4cm}\\
  \subfloat{\label{fg:01d}\includegraphics[height=0.275\textwidth,trim={0px 0px 0px 0px},clip]{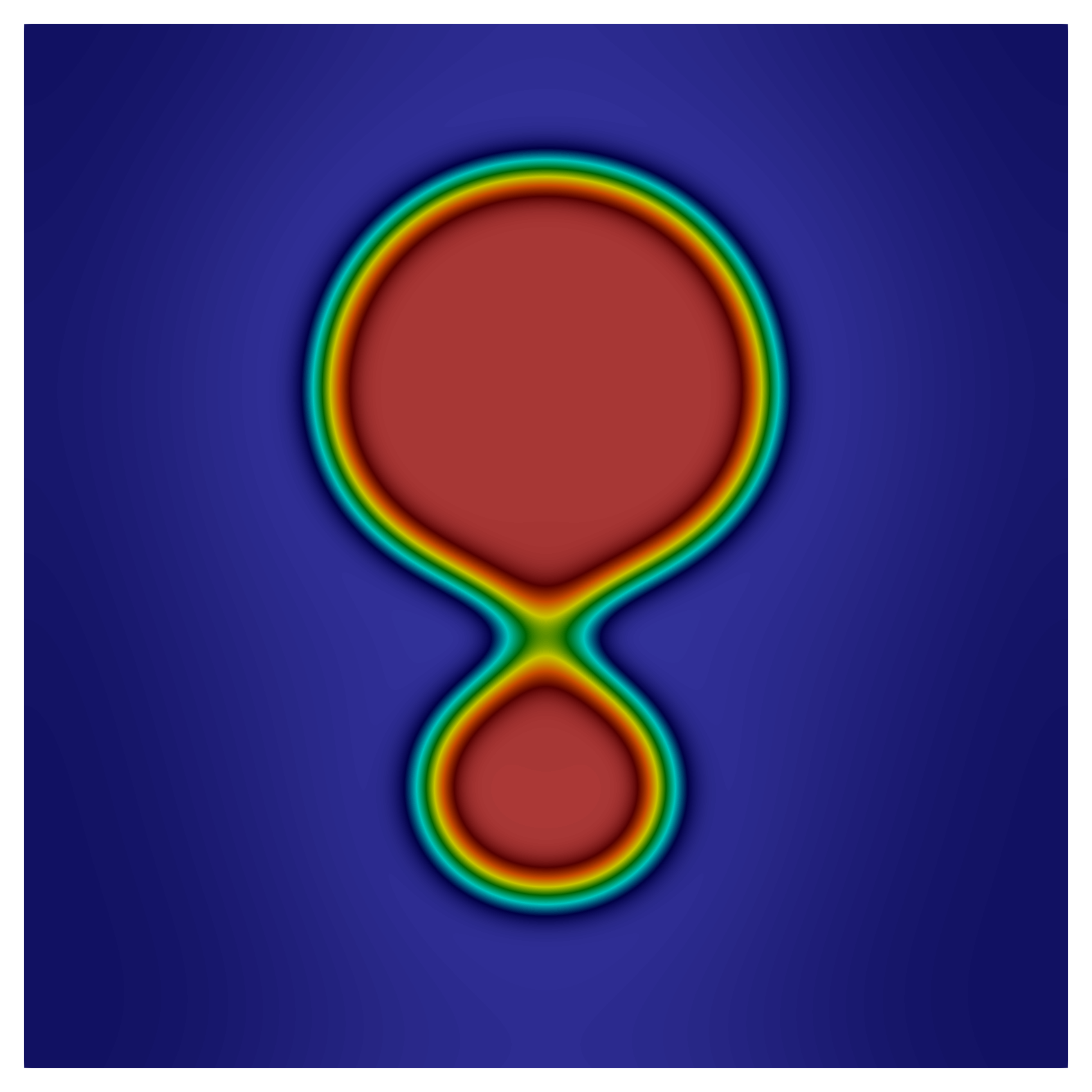}}
  \subfloat{\label{fg:01e}\includegraphics[height=0.275\textwidth,trim={0px 0px 0px 0px},clip]{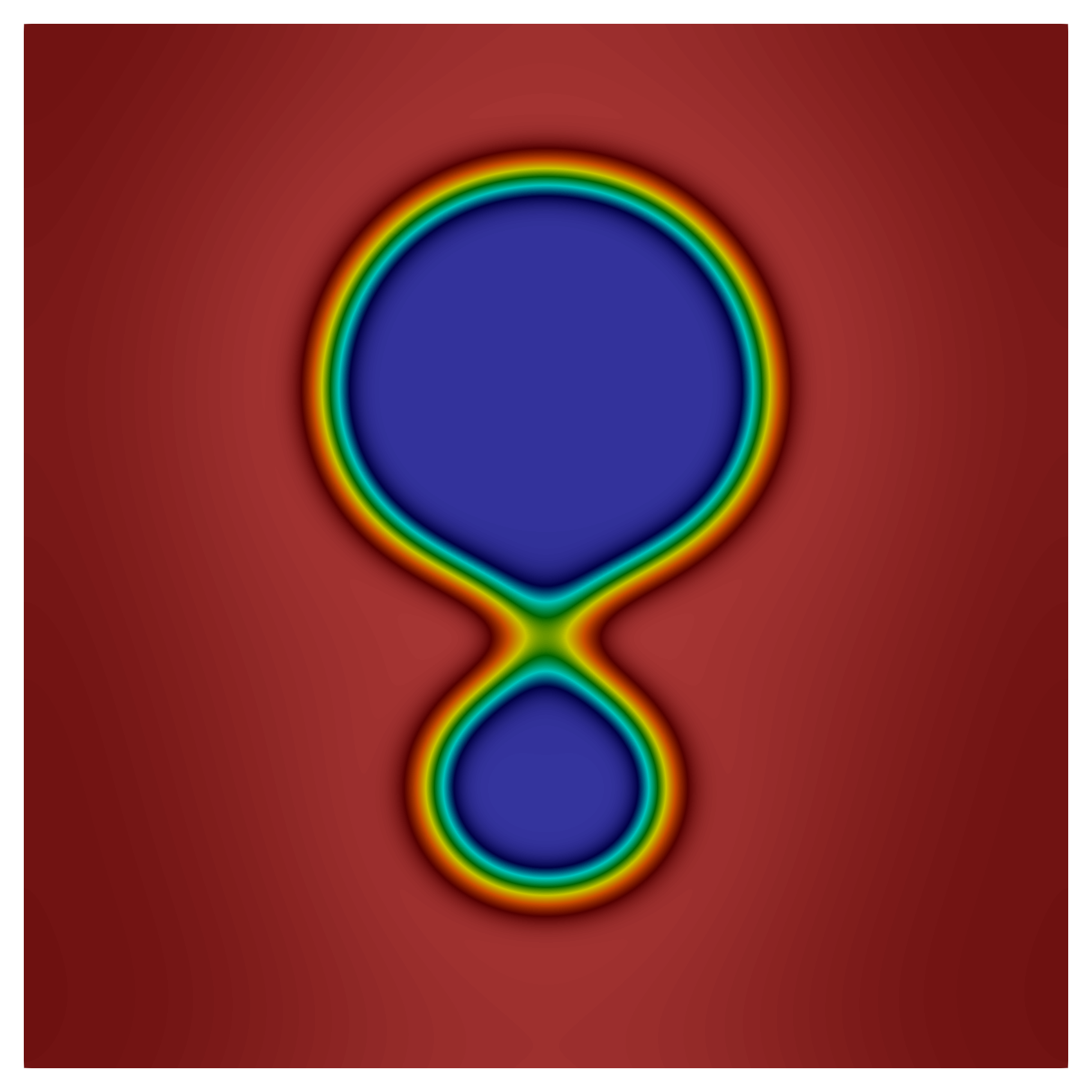}}
  \subfloat{\label{fg:01f}\includegraphics[height=0.275\textwidth,trim={0px 0px 0px 0px},clip]{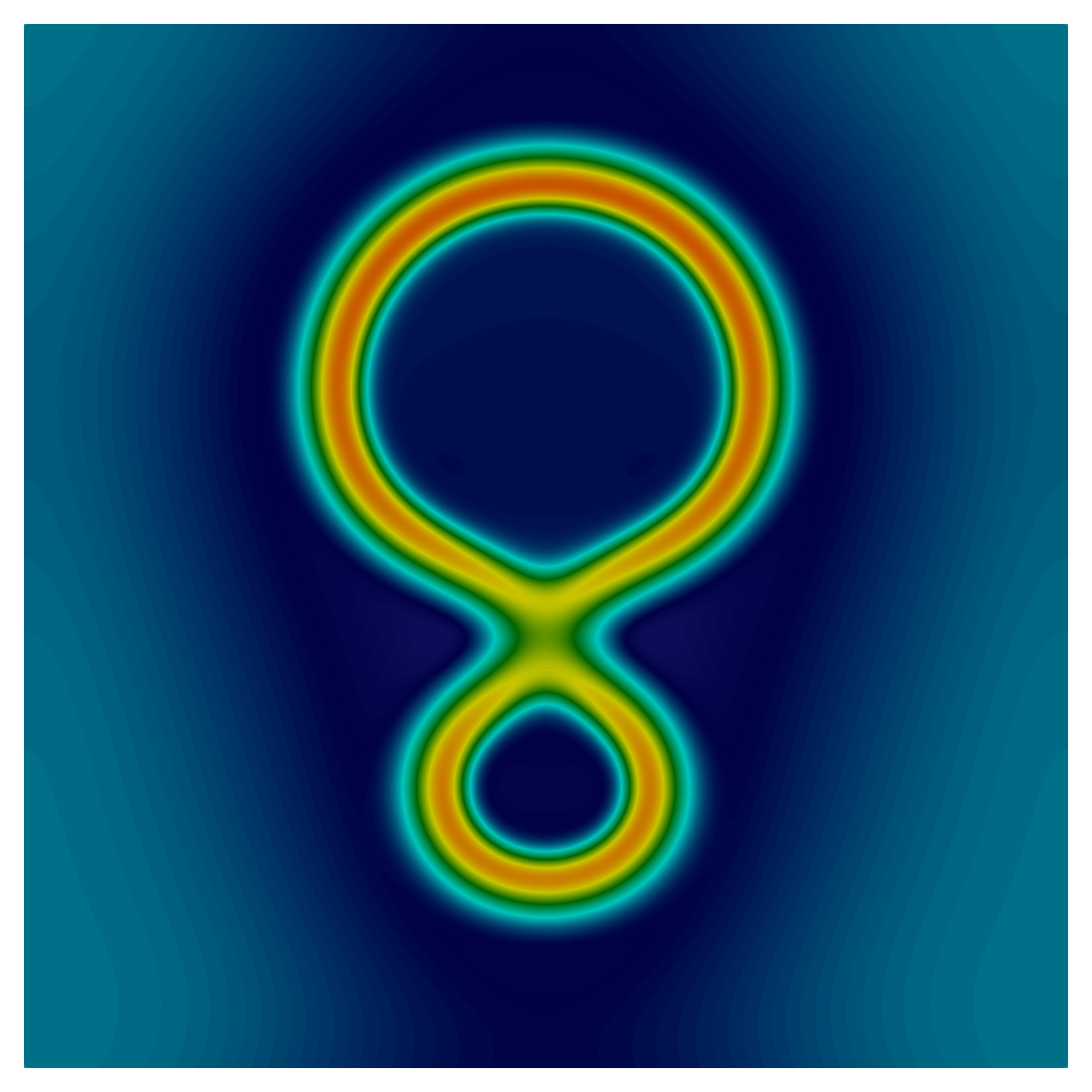}}\vspace{-0.4cm}\\
  \subfloat{\label{fg:01g}\includegraphics[height=0.275\textwidth,trim={0px 0px 0px 0px},clip]{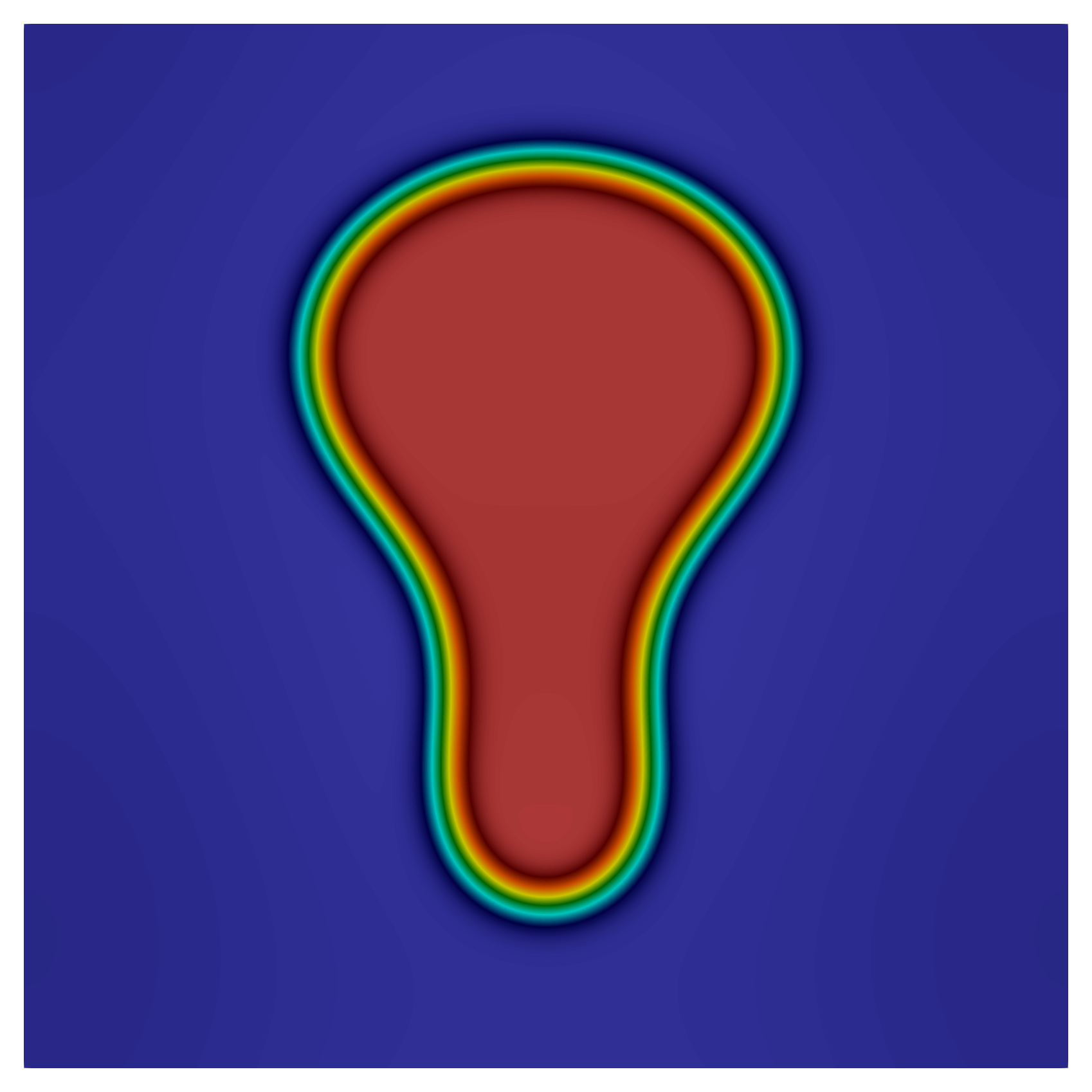}}
  \subfloat{\label{fg:01h}\includegraphics[height=0.275\textwidth,trim={0px 0px 0px 0px},clip]{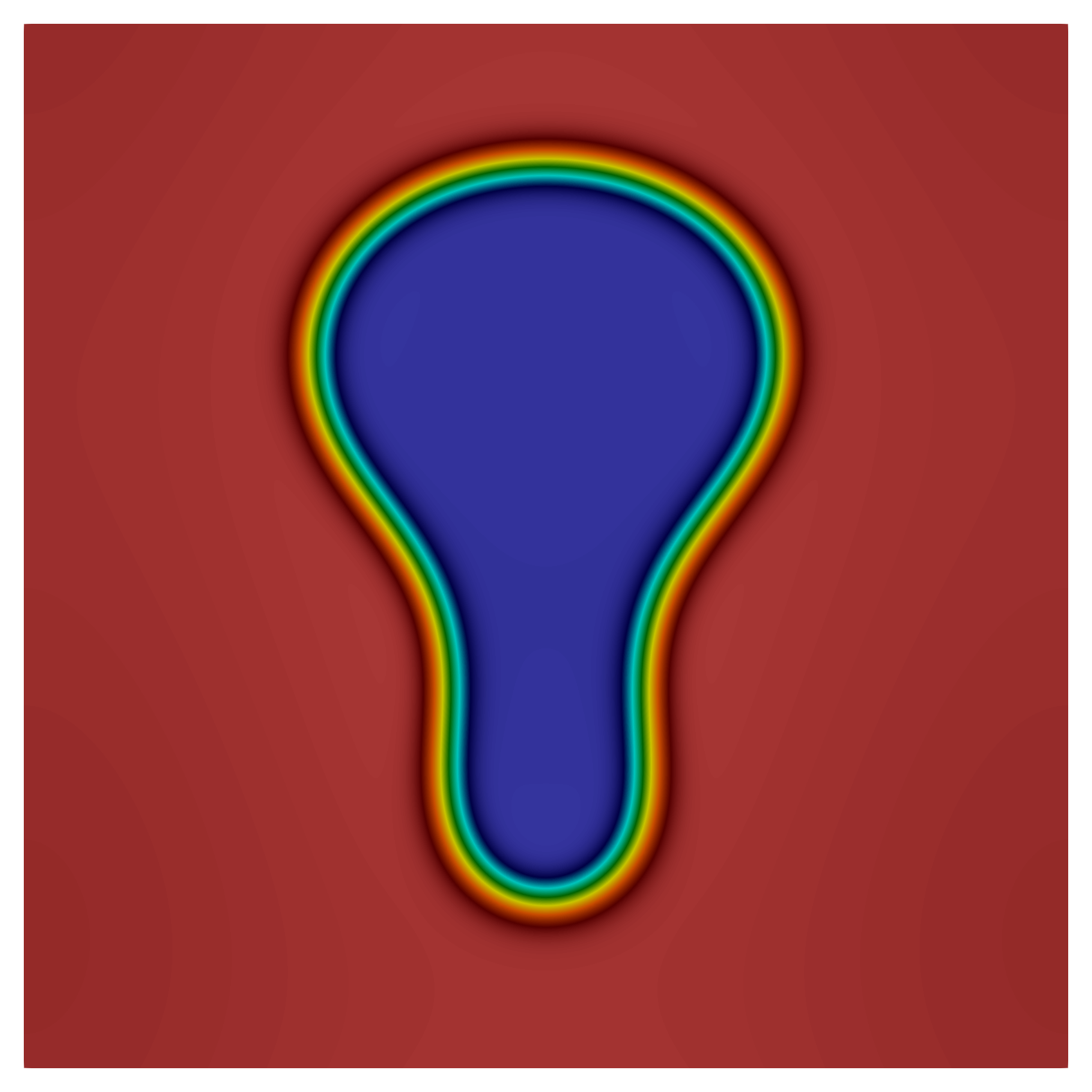}}
  \subfloat{\label{fg:01i}\includegraphics[height=0.275\textwidth,trim={0px 0px 0px 0px},clip]{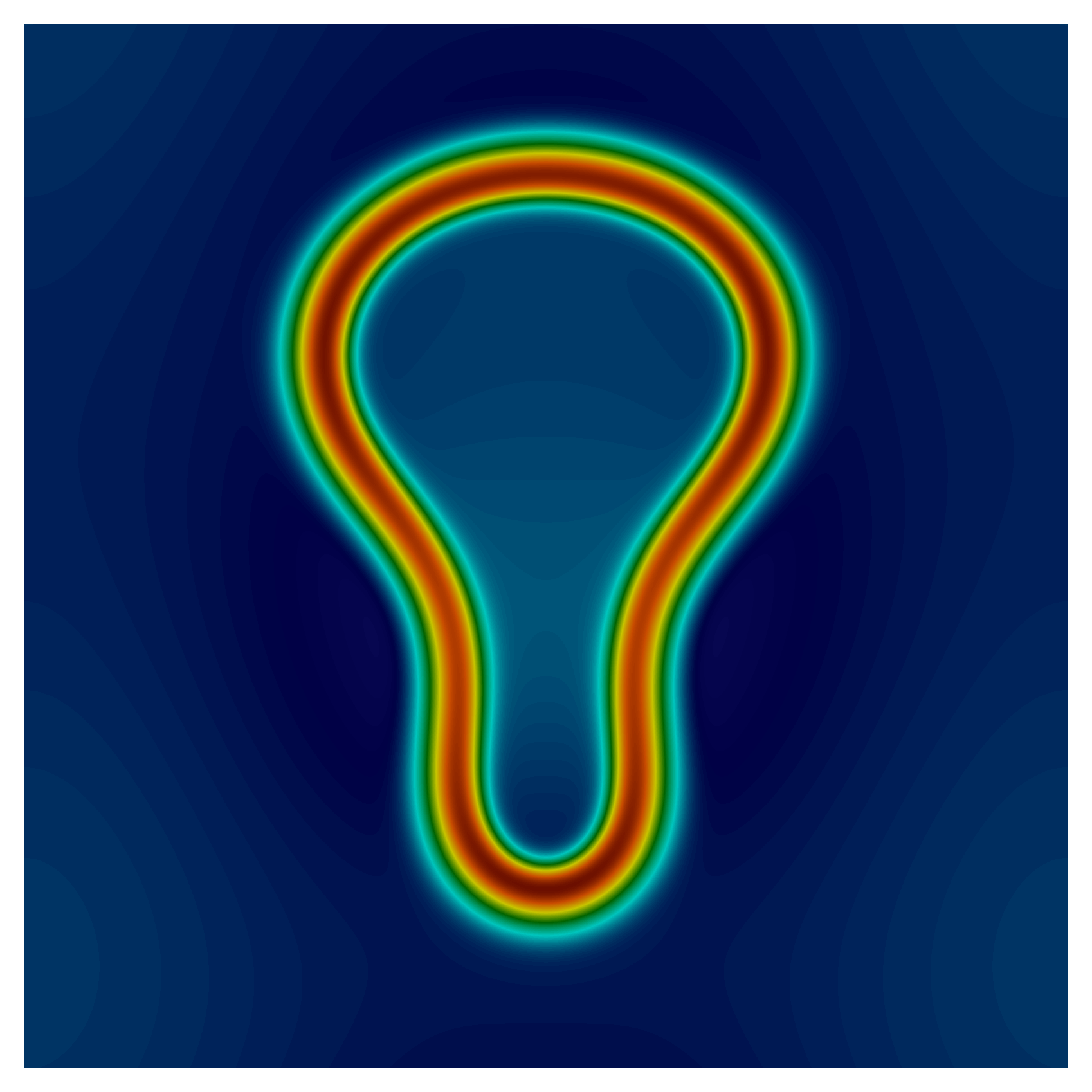}}\vspace{-0.4cm}\\
  \subfloat{\label{fg:01j}\includegraphics[height=0.275\textwidth,trim={0px 0px 0px 0px},clip]{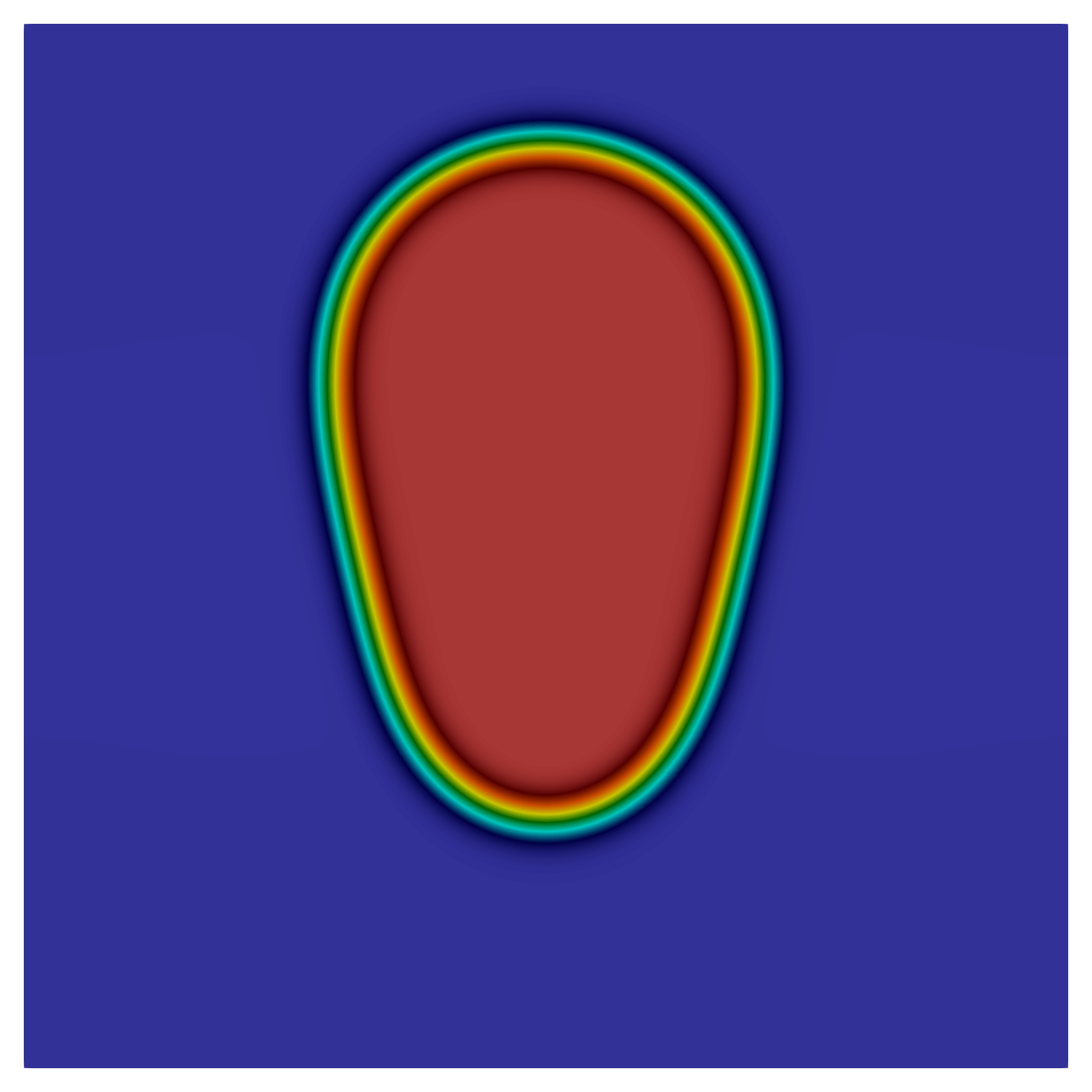}}
  \subfloat{\label{fg:01k}\includegraphics[height=0.275\textwidth,trim={0px 0px 0px 0px},clip]{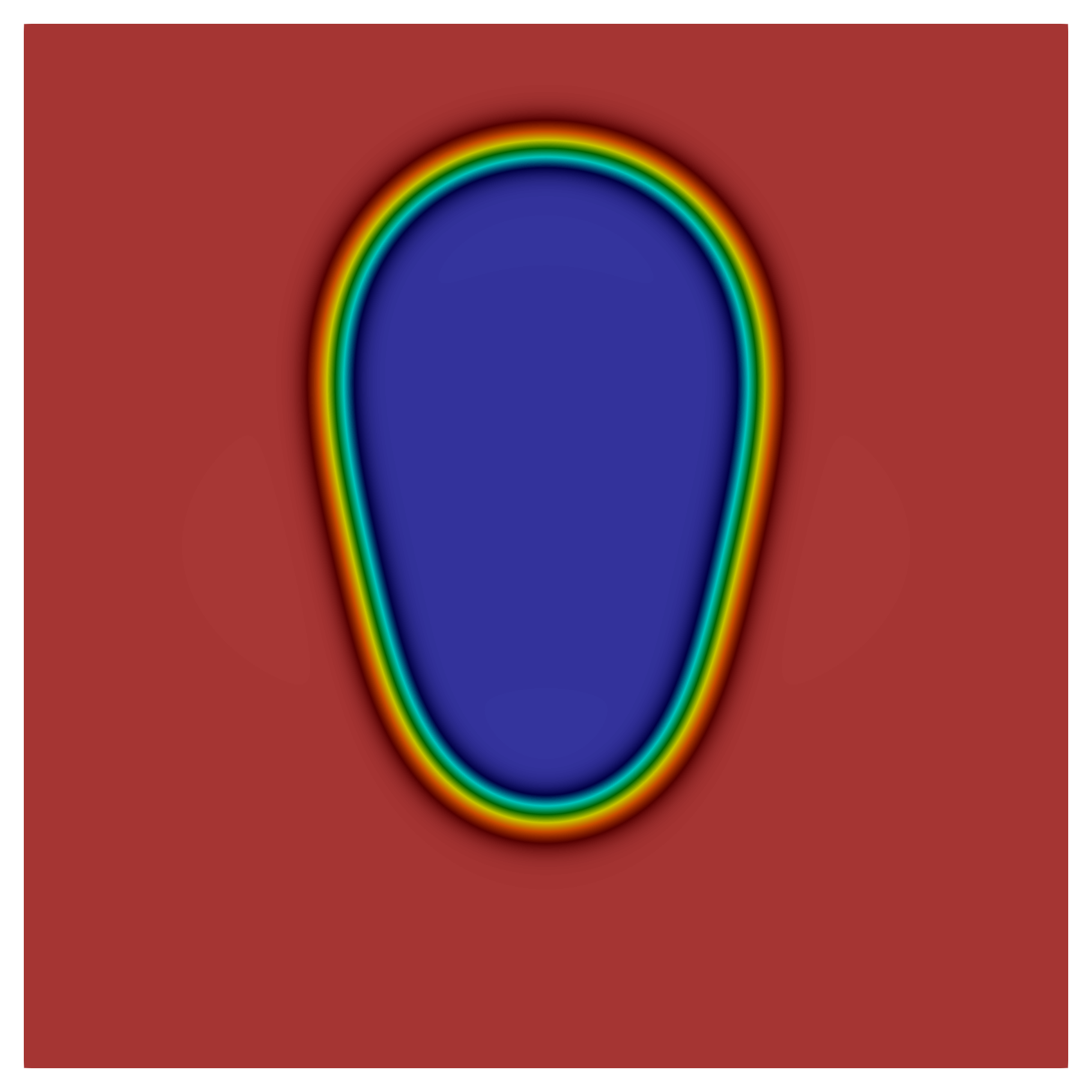}}
  \subfloat{\label{fg:01l}\includegraphics[height=0.275\textwidth,trim={0px 0px 0px 0px},clip]{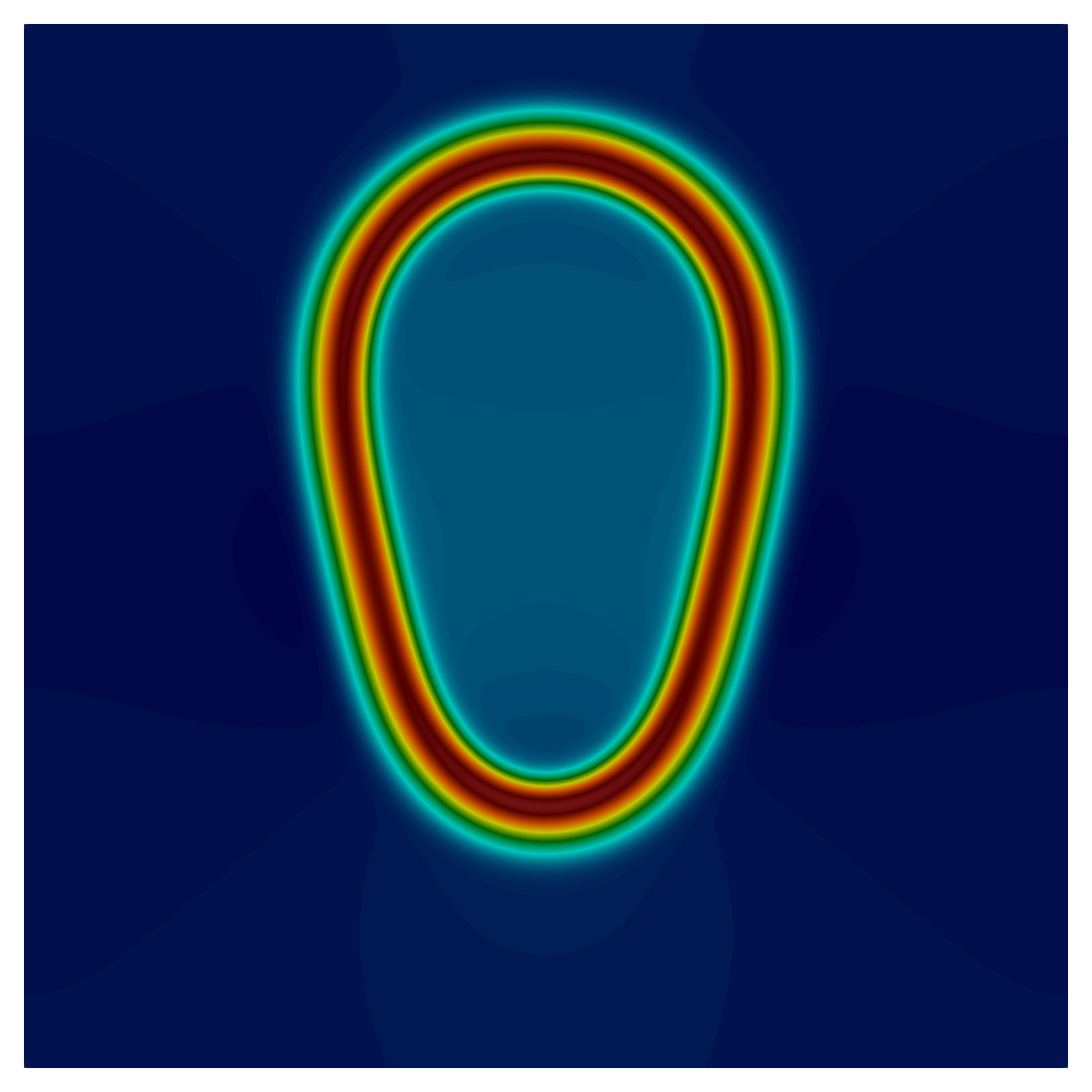}}\vspace{-0.4cm}\\
  \subfloat{\label{fg:01m}\includegraphics[height=0.275\textwidth,trim={0px 0px 0px 0px},clip]{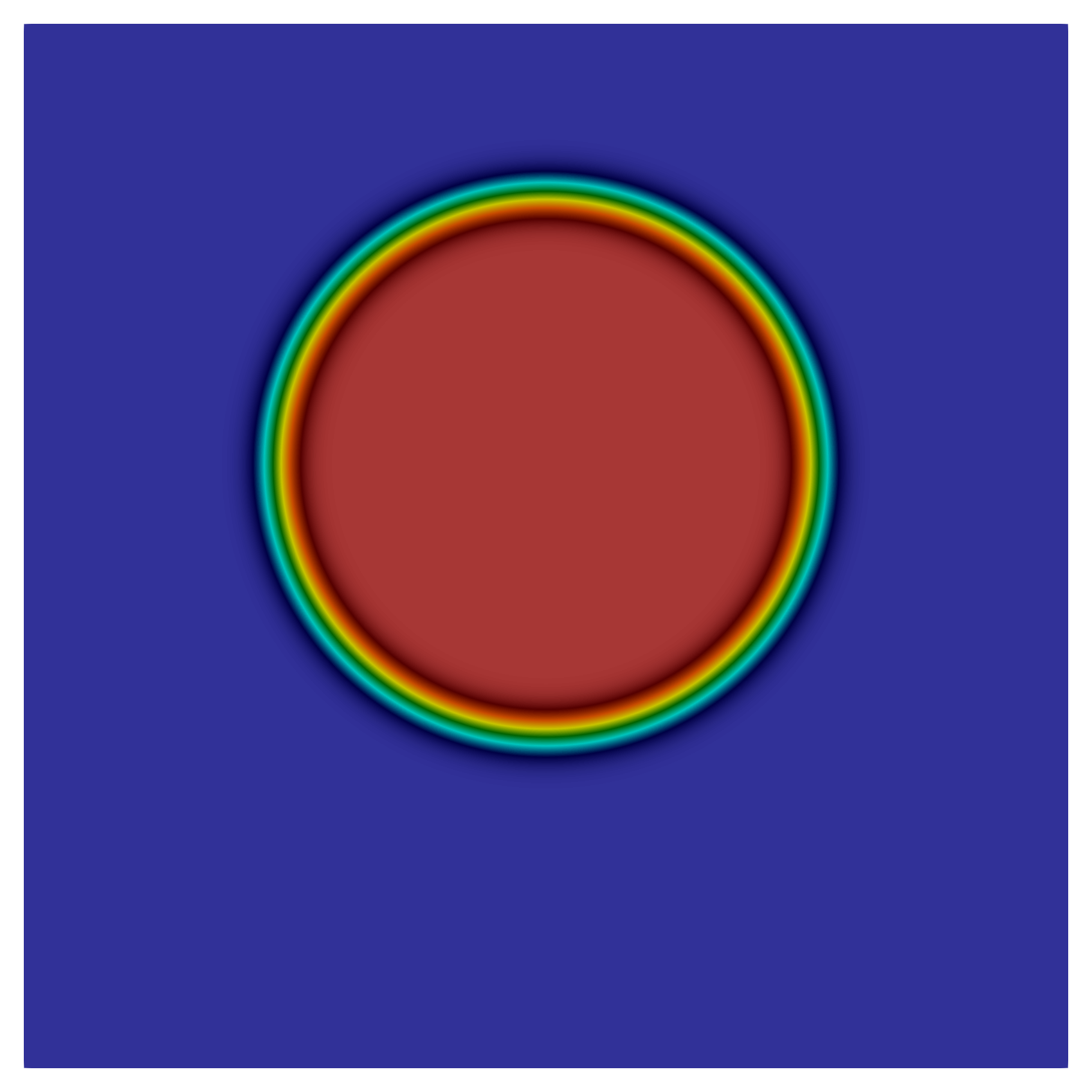}}
  \subfloat{\label{fg:01n}\includegraphics[height=0.275\textwidth,trim={0px 0px 0px 0px},clip]{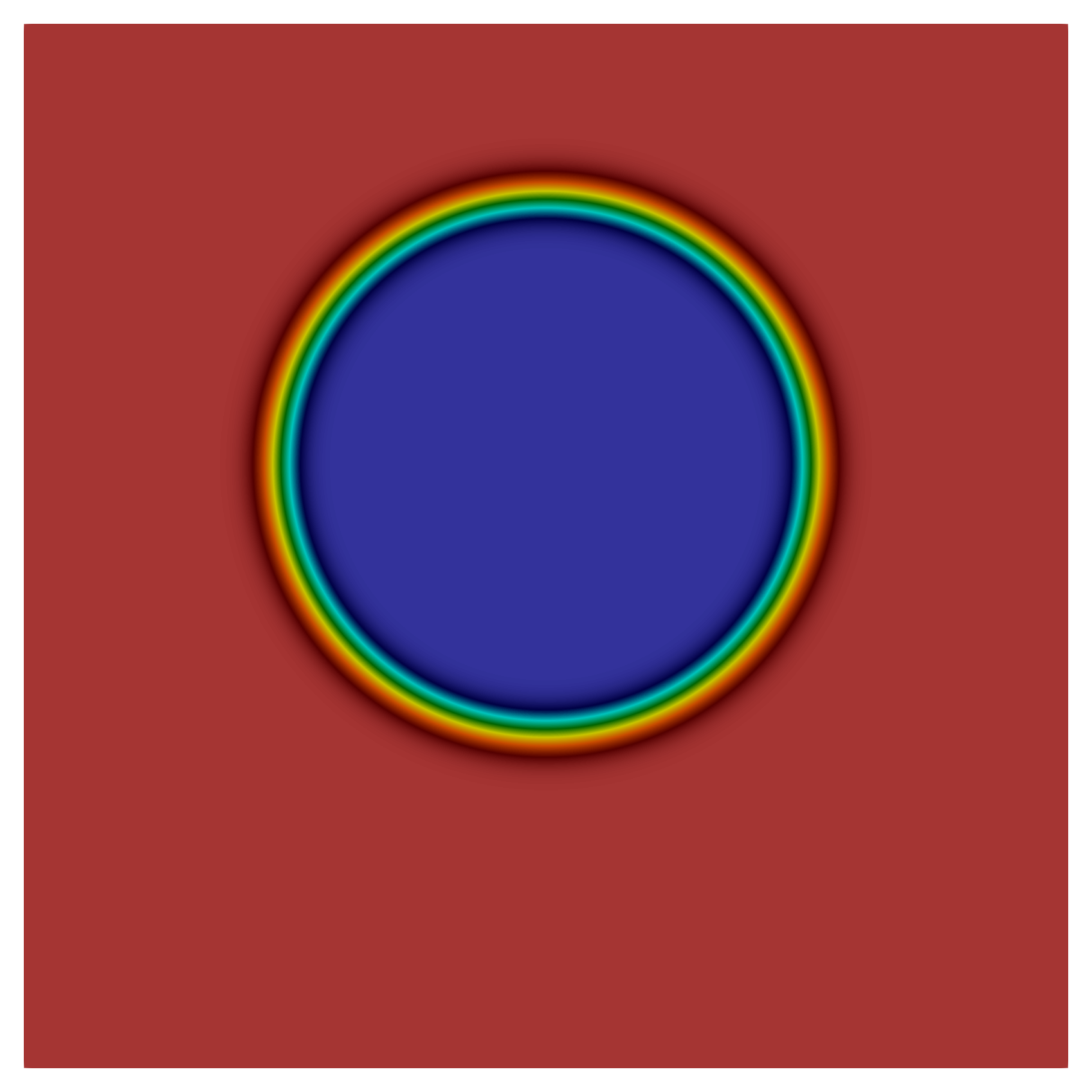}}
  \subfloat{\label{fg:01o}\includegraphics[height=0.275\textwidth,trim={0px 0px 0px 0px},clip]{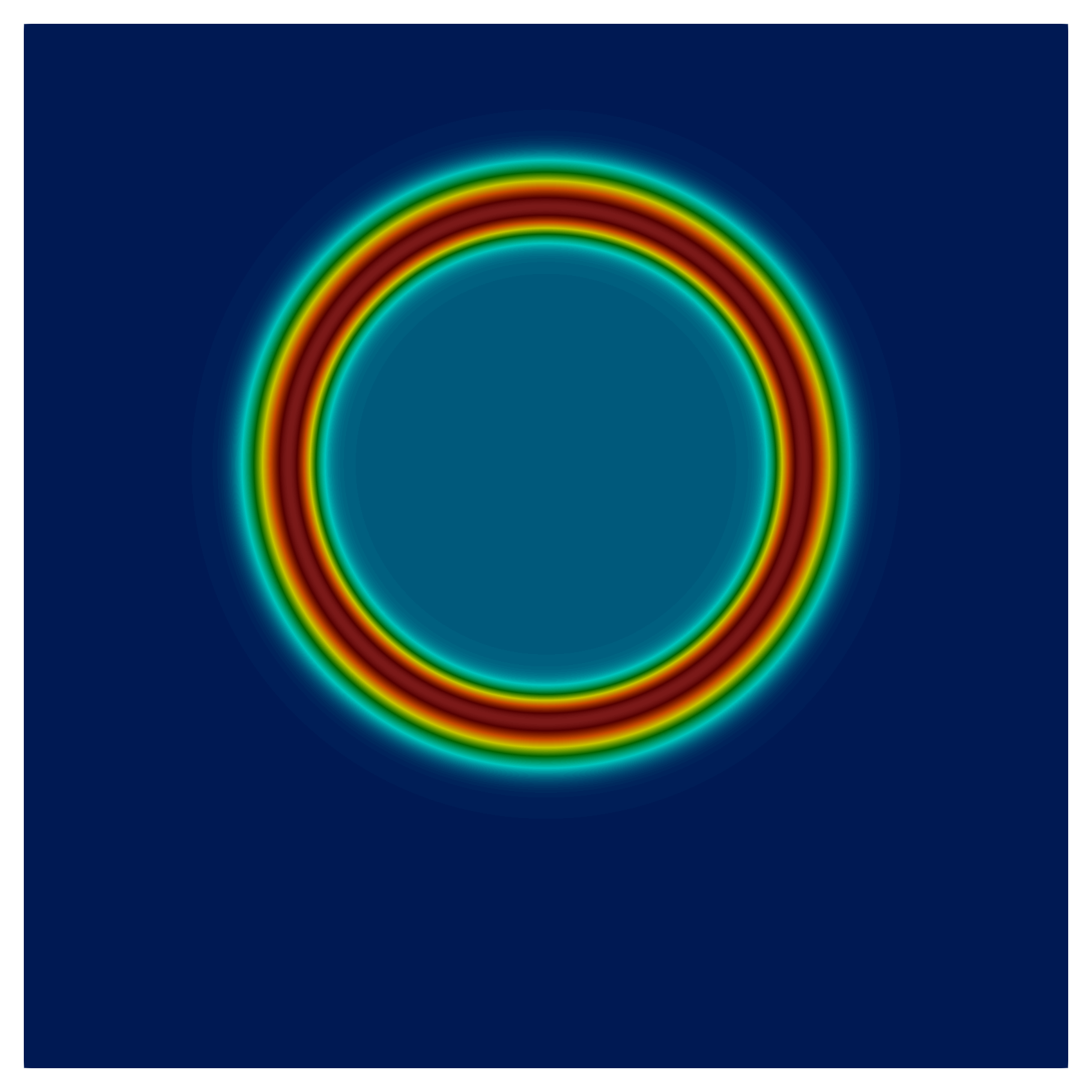}}
  \caption{Phase-field evolution during merging. From left to right ${\varphi}^1$, ${\varphi}^2$, and ${\varphi}^3$. From top to bottom $t=0$, $t=6.00\times10^{-6}$, $t=2.25\times10^{-5}$, $t=7.41\times10^{-5}$, $t=5.96\times10^{-2}$.}
  \label{fg:01}
\end{figure}

We now simulate the interactions between three species where $\calA^1$ and $\calA^2$ represent the reactants, while $\calA^3$ the reaction products. The inclusions (represented by species $1$, $\calA^1$) are embedded in species $2$, $\calA^2$. We express the chemical reaction as
\begin{equation}
\calA^1 + \calA^2 \mathop{\rightharpoonup}^{\bar{k}_+}_{\phantom{\bar{k}_-}} \calA^3
\end{equation}
which takes place at the interface producing the third species, $\calA^3$.

\begin{table}[!htb]
\centering
\caption{Chemical and physical parameters}
\begin{tabular}{c c c}
\hline
Physical parameter & Value & Name \\
\hline
$\psi_0$ [\si{J.m^{-3}}] & $2 \times 10^{7}$ & Energy density\\
$L_0$ [\si{m}] & $10^{-6}$ & Domain length \\
$\vartheta$ [\si{K}] & 1000.0 & Absolute temperature\\
$\vartheta_{c}^{12}$ [\si{K}] & 1100.0 & Critical temperature between phases 1 and 2  \\
$\vartheta_{c}^{13}$ [\si{K}] & 1200.0 & Critical temperature between phases 1 and 3  \\
$\vartheta_{c}^{23}$ [\si{K}] & 1300.0 & Critical temperature between phases 2 and 3  \\
$D$ [\si{m^2.s^{-1}}] &$10^{-20}$& Diffusion coefficient (for all phases) \\
$k^+$ [\si{m^2.s^{-1}}] &$10^{-14}$ & Forward reaction rate \\
$\sigma^{12}$ [\si{J.m^{-2}}] & 0.816& Interfacial energy between phases 1 and 2 \\
$\sigma^{13}$ [\si{J.m^{-2}}] & 0.625& Interfacial energy between phases 1 and 3 \\
$\sigma^{23}$ [\si{J.m^{-2}}] & 0.921& Interfacial energy between phases 2 and 3 \\
$\ell^{12}$ [\si{m}] &$ 1.5 \times 10^{-8}$& Interface thickness between phases 1 and 2 \\
$\ell^{13}$ [\si{m}] &$ 2 \times 10^{-8}$& Interface thickness between phases 1 and 3 \\
$\ell^{23}$ [\si{m}] &$10^{-8}$& Interface thickness between phases 2 and 3 \\
\hline
\end{tabular}
\label{tb:pcparameters}
\end{table}

We state the problem as: find $\bfvarphi$ satisfying~\eqref{eq:gov.eq.dimless} given~\eqref{eq:gov.eq.bvp} subject to periodic boundary conditions up to the fourth derivative of $\bfvarphi$ with respect to $\bfx$ in a square open region $\calD=(0,1)\times(0,1)$. We discretize the resulting system of partial differential equations using  PetIGA~\cite{ Dal16}, a high-performance isogeometric analysis framework. We solve this system of equations in their primal form using a~$256\times256$ element mesh of a polynomial degree $4$ and continuity $3$. The initial and boundary conditions are
\begin{equation}
\left\{\,
\begin{aligned}
&h=0.035& \\
&\delta^1(\bfx,0) = -0.76\left(0.5\tanh\left(\frac{(x-0.5)^2+ (y-0.65)^2}{h(h+0.2)} \right)+ 0.5\right)+0.31\\
&\delta^2(\bfx,0) = -0.76\left(0.5\tanh\left(\frac{(x-0.5)^2+ (y-0.29)^2}{h(h+0.1)} \right)+ 0.5\right)+0.31\\
&\varphi^1(\bfx,0)=1+\delta_1+\delta_2,&\text{in}\,\calD,\\
&\varphi^2(\bfx,0)=0.999-\varphi^1,&\text{in}\,\calD,\\
&\varphi^3(\bfx,0)=1-\varphi^1-\varphi^2,&\text{in}\,\calD,\\
&\text{subject to periodic boundary conditions}&\text{on}\,\,\partial\calD\times(0,T),
\end{aligned}
\right.
\end{equation}
and the three subfigures on top of Figure~\ref{fg:01} depict this initial condition.

Table~\ref{tb:pcparameters} summarizes the dimensional parameters used to obtain the dimensionless parameters in~\eqref{eq:dimensionless.parameters.D} and~\eqref{eq:dimensionless.parameters}. The diffusion matrix for each entry $\alpha$ and $\beta$ reads
\begin{equation}\label{eq:dimensionless.parameters.D}
  \bar{\bfD}^{\alpha\beta}=1\times10^{4}\varphi^{\alpha}(\delta^{\alpha\beta} - \varphi^{\beta})
  \begin{bmatrix}
    1 & 1 & 1 \\
    0 & 1 & 1 \\
    0 & 0 & 1
  \end{bmatrix}\qquad\forall\,1\le\alpha,\beta\le{n}.
\end{equation}
Next, for clarity, we represent $\alpha$ and $\beta$ as matrix-columns and -rows indices, which render the remaining dimensionless parameters as follows.
\begin{equation}\label{eq:dimensionless.parameters}
  \begin{gathered}
    \bar{\sigma}^{\alpha\beta}\bar{\ell}^{\alpha\beta}=-10^{-4}
    \begin{bmatrix}
      0 & 6.121 & 6.250 \\
      6.121 & 0 & 4.605 \\
      6.250 & 4.605 & 0
    \end{bmatrix},\quad
    \upsilon^{\alpha\beta}=
    \begin{bmatrix}
      1 & 1 & 0
    \end{bmatrix},\quad
    \varpi^{\alpha\beta}=
    \begin{bmatrix}
      0 & 0 & 1
    \end{bmatrix},\quad
    \bar{k}_+=0.01,
  \end{gathered}
\end{equation}
where we choose $D_0 = D$ and $\ell_0 = \ell^{23}$ as the reference diffusion coefficient and interface thickness of a reference species, respectively.

Here, the configurational tractions drive the interfacial motion in this multicomponent system undergoing reactions. We express the configurational traction along a level curve $\calL^\alpha_*$, upon which $\varphi^\alpha=\varphi^\alpha_*$. We then introduce the normal and tangential coordinates $n^\alpha$ and $m^\alpha$ on $\calL^\alpha_*$, with unit vectors $\bfnu^\alpha$ and $\bftau^\alpha$ defined such that
\begin{equation}
  \Grad\varphi^\alpha=|\Grad\varphi^\alpha|\bfnu^\alpha,\qquad\qquad|\bfnu^\alpha|=1,
\end{equation}
augmented by a sign convention which ensures that rotating $\bftau^\alpha$ clockwise by $\pi/2$ yields $\bfnu^\alpha$. In reckoning the relative configurational stress in a $\{n^\alpha,m^\alpha\}$-frame, we arrive at
\begin{equation}\label{eq:configurational.stress.surf}
  \bfC_\sigma=\zeta\id-\sum_{\alpha=1}^n|\Grad\varphi^\alpha|\bfnu^\alpha\otimes\bfxi^\alpha_\sigma,
\end{equation}
with
$\zeta\coloneqq\varrho\left(\psi-\sum_{\alpha=1}^{n}\mu^\alpha\varphi^\alpha\right)$,
see~\eqref{eq:zeta} in Appendix~\ref{sc:configurationalstresses}. We can now specialize~\eqref{eq:configurational.stress.surf} with a free-energy of the form
\begin{align}
\hat\psi(\bfvarphi,\Grad\bfvarphi)&=f(\varphi^\alpha)+\frac{1}{2}\sum_{\alpha=1}^n\sum_{\beta=1}^n\Gamma^{\alpha\beta}\Grad\varphi^\alpha\cdot\Grad\varphi^\beta, \nonumber \\
&=f(\varphi^\alpha)+\frac{1}{2}\sum_{\alpha=1}^n\sum_{\beta=1}^n\Gamma^{\alpha\beta}|\Grad\varphi^\alpha||\Grad\varphi^\beta| \bfnu^\alpha\cdot\bfnu^\beta,
\end{align}
which renders the following relative configurational stress
\begin{equation}\label{eq:configurational.stress.specialized}
\bfC_\sigma=\zeta\id+\sum_{\alpha=1}^n\left\{|\Grad\varphi^\alpha|\bfnu^\alpha\otimes\left(\sum_{\beta=1}^n(\Gamma^{\alpha\beta}-\Gamma^{\sigma\beta})|\Grad\varphi^\beta|\bfnu^\beta\right)\right\}.
\end{equation}
Thus, the configurational tractions $\bfC_\sigma\bfnu^\alpha$ are
\begin{align}\label{eq:configurational.traction.on.L}
\bfC_\sigma\bfnu^\alpha&=\left\{\zeta+\sum_{\hat{\alpha}=1}^n\sum_{\hat{\beta}=1}^n\left((\Gamma^{\hat{\alpha}\hat{\beta}}-\Gamma^{\sigma\hat{\beta}})|\Grad\varphi^{\hat{\alpha}}||\Grad\varphi^{\hat{\beta}}|\bfnu^{\hat{\alpha}}\otimes\bfnu^{\hat{\beta}}\right)\right\}\bfnu^\alpha\nonumber\\
&=\zeta\bfnu^\alpha+\sum_{\hat{\alpha}=1}^n\left((\Gamma^{\hat{\alpha}\alpha}-\Gamma^{\sigma\alpha})|\Grad\varphi^{\hat{\alpha}}||\Grad\varphi^{\alpha}|\bfnu^{\hat{\alpha}}\right).
\end{align}
In the simulations, we compute the relative physical and chemical quantities, such as the relative chemical potential, mass fluxes, microstresses, and byproducts, by setting the reaction product species, that is, $\calA^3$ as the reference phase field. This simulation shows that the configurational fields can describe the behavior of the phase evolution. However, this initial work does not exploit this tool exhaustively nor comprehensively.

Figure~\ref{fg:01} depicts the merging process of two circular inclusions of distinct size into a single one. The figure spans from the early stages until the merged inclusion becomes stationary. From left to right, we depict phases $\varphi^1$, $\varphi^2$, and $\varphi^3$, while from top to bottom, the evolution of the three phases for the dimensionless times $t=0$, $6.00\times10^{-6}$, $2.25\times10^{-5}$, $7.41\times10^{-5}$, and $5.96\times10^{-2}$.

\begin{figure}[!htb]
  \includegraphics[height=0.975\textwidth]{cn2}
  \caption{Vertical component of the configurational traction $\bfC_3\bfnu^2$ along $x_1=0.5$.}
  \label{fg:04}
\end{figure}

\begin{figure}[!htb]
  \includegraphics[height=0.975\textwidth]{cn3}
  \caption{Vertical component of the configurational traction $\bfC_3\bfnu^3$ along $x_1=0.5$.}
  \label{fg:05}
\end{figure}

\begin{figure}[!htb]
  \centering
  \subfloat{\label{fg:06a}\includegraphics[height=0.25\textwidth,clip]{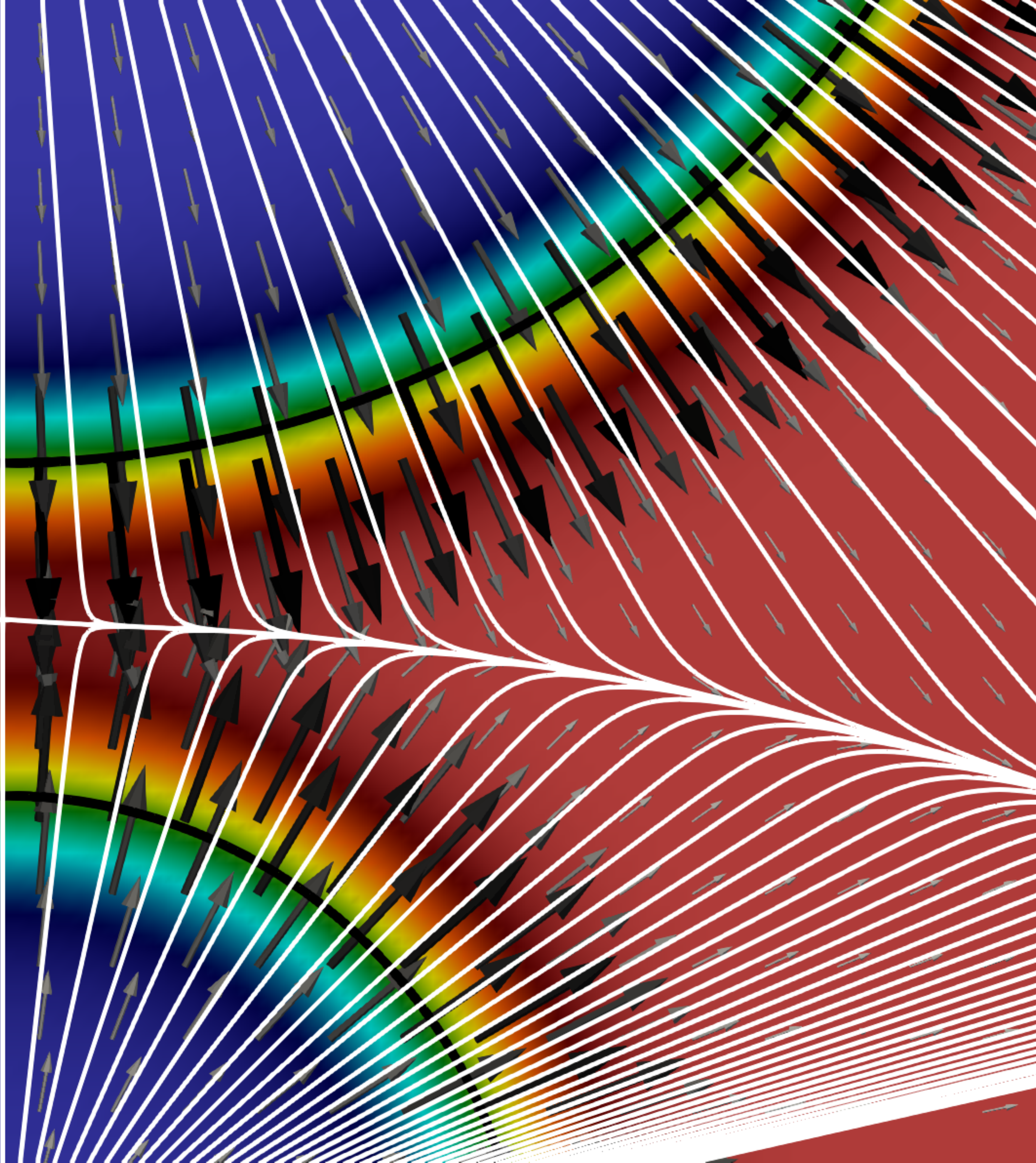}}  \hspace{0.1cm}
  \subfloat{\label{fg:06c}\includegraphics[height=0.25\textwidth,clip]{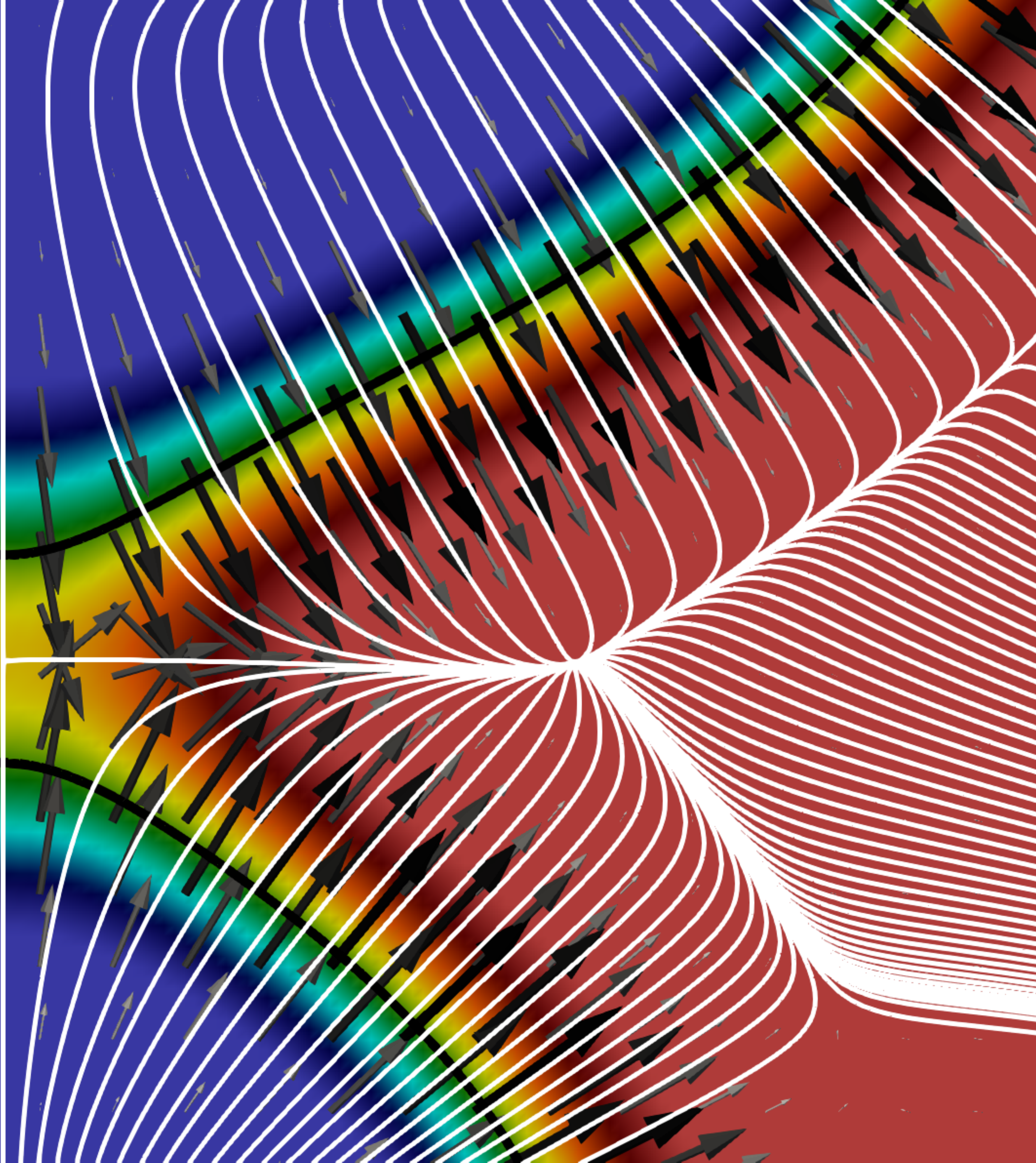}}  \\
  \subfloat{\label{fg:06d}\includegraphics[height=0.25\textwidth,clip]{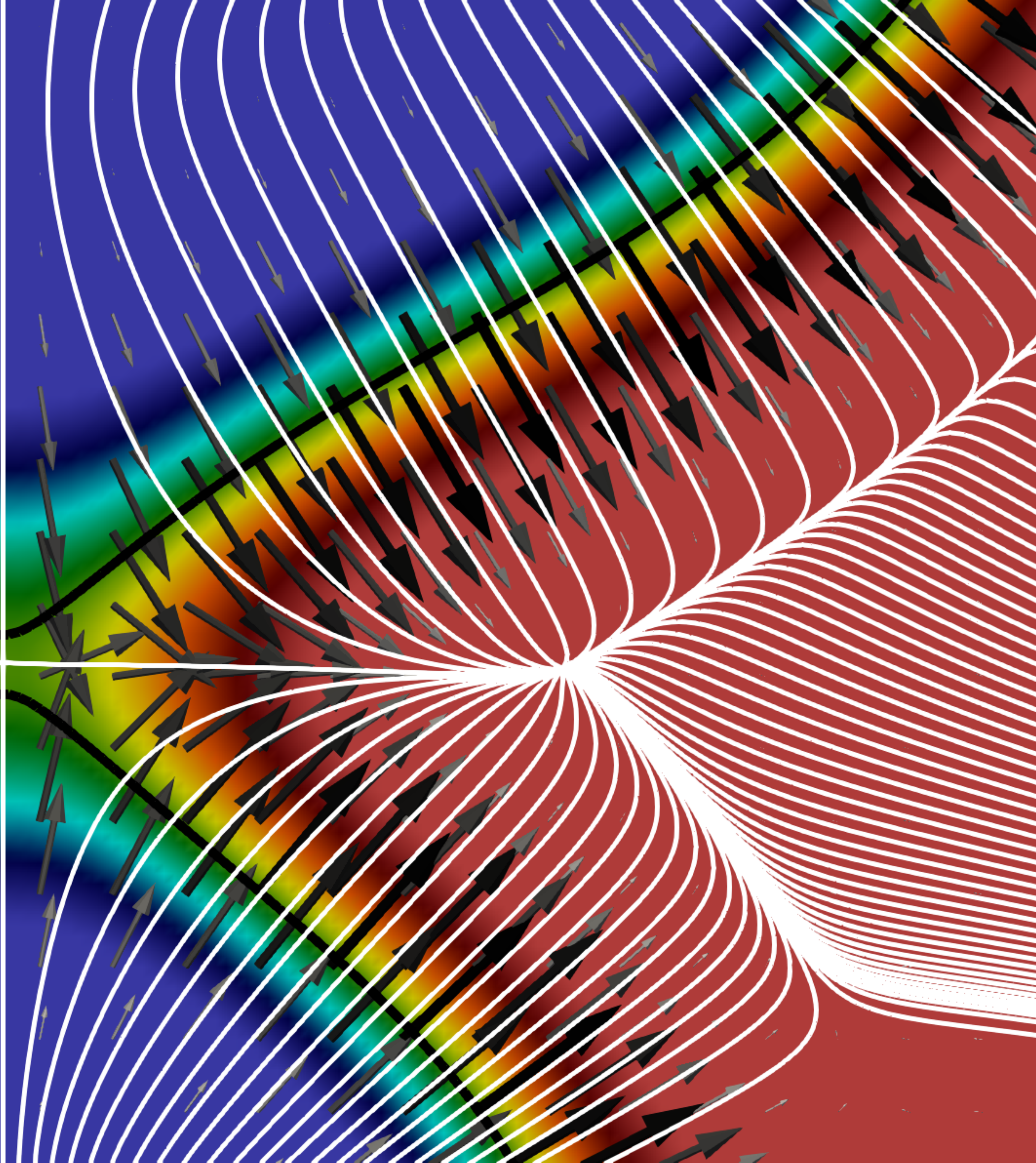}}  \hspace{0.1cm}
  \subfloat{\label{fg:06e}\includegraphics[height=0.25\textwidth,clip]{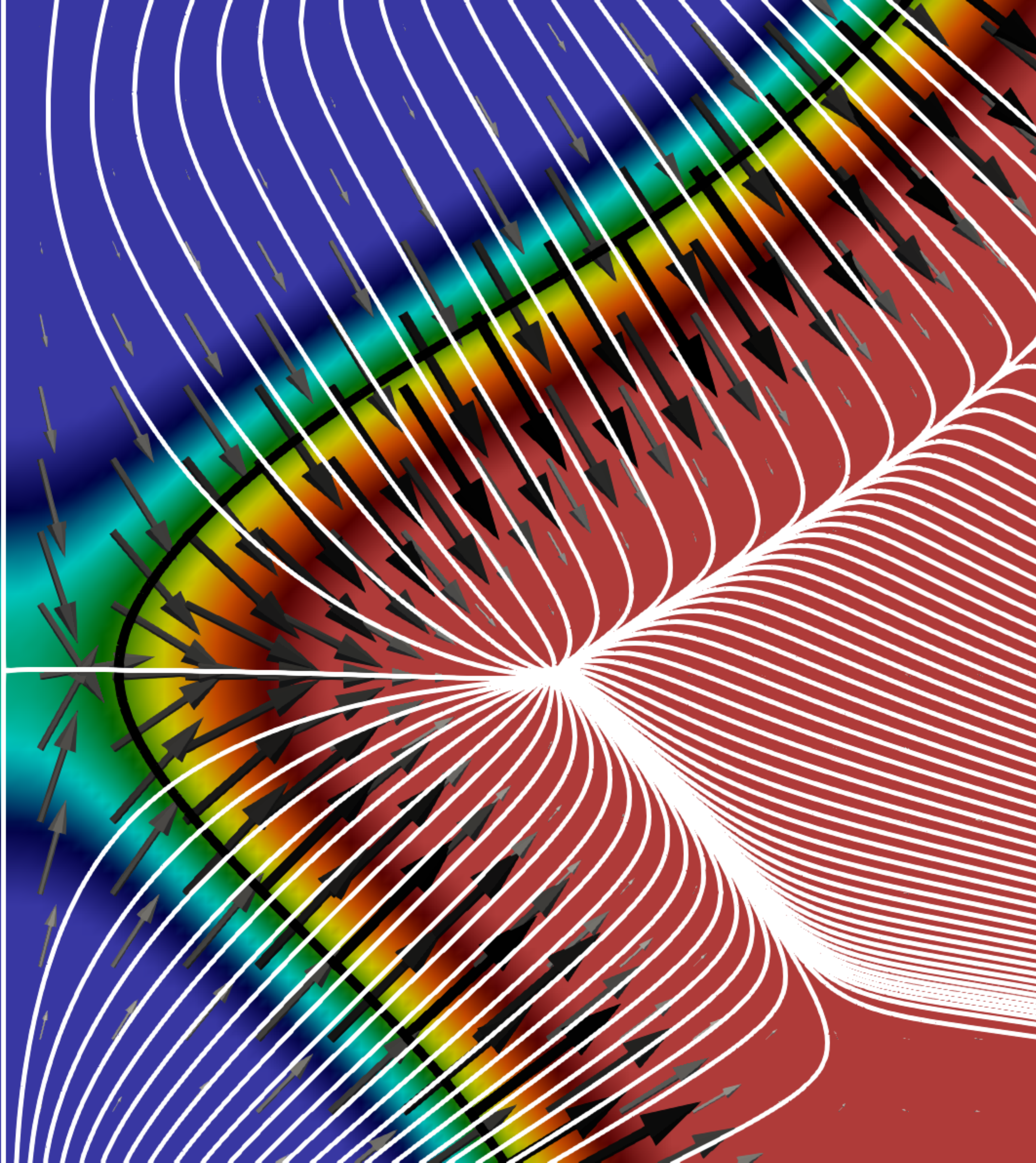}}
  \caption{Node sinks triggering the merging $\bfC_3\bfnu^2$. From left to right, the relative configurational traction $\bfC_3\bfnu^2$, streamlines of this traction (white), $\calL_{0.5}^2$ (black) on top of the phase field $\varphi^1$ at the dimensionless times $t=0$, $5.39\times10^{-6}$, $6.00\times10^{-6}$, $6.57\times10^{-6}$, and $2.25\times10^{-5}$.}
  \label{fg:06}
\end{figure}

% To compute the relative configurational traction, we choose the level curves $\calL_{0.5}^2$ and $\calL_{2e-3}^3$ with their corresponding outer unit normals, being $\calA^3$ the reference species. Figure~\ref{fg:02} shows the evolution of the relative configurational traction $\bfC_3\bfnu^2$ on the level curve $\calL_{0.5}^2$. The relative configurational traction $\bfC_3\bfnu^2$ is reckoned with the outer unit normal $\bfnu^2\coloneqq\Grad\varphi^2/|\Grad\varphi^2|$. $\bfnu^2$ is the outer normal to the inclusions, formed by the first and second phases. Likewise, in Figure~\ref{fg:03}, we present the evolution of the relative configurational traction $\bfC_3\bfnu^3$ on the level curve $\calL_{2e-3}^3$. The relative configurational traction $\bfC_3\bfnu^3$ is reckoned with the outer unit normal $\bfnu^3\coloneqq-\Grad\varphi^3/|\Grad\varphi^3|$. $\bfnu^3$ is the outer normal of the ring-like structures, formed by the third phase. In Figures~\ref{fg:02} and~\ref{fg:03}, from left to right, the evolution of the relative configurational tractions is presented for the dimensionless times $t=0$, $5.39\times10^{-6}$, $6.00\times10^{-6}$, $6.57\times10^{-6}$, $2.25\times10^{-5}$, $7.41\times10^{-5}$, and $5.96\times10^{-2}$.

Figures~\ref{fg:04} and~\ref{fg:05}, respectively, present $\bfe_2\cdot\bfC_3\bfnu^2(x_1=0.5,x_2)$ and $\bfe_2\cdot\bfC_3\bfnu^3(x_1=0.5,x_2)$ on the left panel, and $\bfe_2\cdot\bfC_3\bfnu^2(\calD)$ and $\bfe_2\cdot\bfC_3\bfnu^3(\calD)$ on the right panel. That is, the left panels display the profile of the relative configurational traction along $x_2$, while the right panels display the vertical component of the relative configurational traction on the whole domain. These figures show the $x_2$ axis in red. From top to bottom, we present these configurational fields at the dimensionless times $t=0$, $5.39\times10^{-6}$, $6.00\times10^{-6}$, $6.57\times10^{-6}$, $2.25\times10^{-5}$, $7.41\times10^{-5}$, and $5.96\times10^{-2}$.  Figure~\ref{fg:04} shows that configurational tractions between the inclusions have opposite directions pushing against one another. As the inclusions approach each other, the configurational traction profiles become antisymmetric in the region where the merging takes place (second plot, from top to bottom, in Figure~\ref{fg:04}). In this region, the ridge and the valley propagate towards each other until the interfaces merge. Later, the configurational tractions annihilate one another (third plot, from top to bottom, in Figure~\ref{fg:04}). The third species appears as the chemical reaction takes place. Figure~\ref{fg:05} shows how the relative configurational traction $\bfC_3\bfnu^3$ pushes apart the boundaries of the double ring, formed by this species. This traction drives the growth of the area encircled by the double ring, which occurs at the expense of the other two species through the chemical reaction. Figure~\ref{fg:05} (second plot, from top to bottom) shows the tractions on each ring as they push against each other, which favours merging. At later stages, a single ring-like structure remains, formed by the product species. This ring lies in between the interface formed by the reactant species. Consequently, the process reaches a semblance of a steady-state when the product species obstructs further the chemical reactions.

Figure~\ref{fg:06} presents a snapshot sequence detailing the merging process from left to right and top to bottom. We use the relative configurational traction $\bfC_3\bfnu^2$, their streamlines (white), $\calL_{0.5}^2$ (black) on top of the phase field $\varphi^1$ at the dimensionless times $t=0$, $5.39\times10^{-6}$, $6.00\times10^{-6}$, $6.57\times10^{-6}$, and $2.25\times10^{-5}$ to exemplify this evolution. In these snapshots, we show the configurational traction $\bfC_3\bfnu^2$ with black arrows. Before the merging occurs, two node sinks arise, see Figure~\ref{fg:06c}. These node sinks pull the phase $\varphi^2$ initiating the merging process. Soon after the node sinks are formed, Figure~\ref{fg:06d}, phase $\varphi^2$ migrates and leaves a `bridge' between the inclusions. This `bridge' is formed by phase $\varphi^1$. After, the merging of the level curve $\calL^1_{0.5}$ (black line) occurs, see Figure~\ref{fg:06e}.

\section{Final remarks}

In this work, we present a continuum framework to model phase separation processes such as spinodal decomposition during cooling as a result of uphill diffusion. These phases are composed of solid solutions of minerals and diffuse at different rates. In this first attempt to model solid diffusion and chemical reactions between rock minerals, we neglect deformation and heat transfer. To this end, we derive a thermodynamically consistent continuum theory for the multicomponent Cahn--Hilliard equations while accounting for multiple chemical reactions. We consider multiple balances of microforces augmented by multiple constituent content balance equations within an extended Larch{\'e}--Cahn framework. Moreover, we derive a configurational balance that includes all the associated configurational fields in agreement with the Larch{\'e}--Cahn framework. In a simple simulation, we depict the role of the configurational tractions during the merging process coupled with a chemical reaction. Last, in upcoming works, we plan to model the contributions of deformation in the thermodynamic pressure arising from chemical processes such as mass transport, chemical reactions, and interfacial effects.

\section{Acknowledgments}

We are indebted to Professor Eliot Fried. We had many exhaustive discussions in which he gave us valuable ideas, constructive comments, and encouragement. This publication was made possible in part by the CSIRO Professorial Chair in Computational Geoscience at Curtin University and the Deep Earth Imaging Enterprise Future Science Platforms of the Commonwealth Scientific Industrial Research Organisation, CSIRO, of Australia. The European Union's Horizon 2020 Research and Innovation Program of the Marie Sk\l{}odowska-Curie grant agreement No. 777778, and the Mega-grant of the Russian Federation Government (N 14.Y26.31.0013) provided additional support. Lastly, we acknowledge the support provided at Curtin University by The Institute for Geoscience Research (TIGeR) and by the Curtin Institute for Computation.

\appendix

\section{Thermodynamically consistent continuum theory for the multicomponent Cahn--Hilliard equations}\label{ap:all}

\subsection{Larch{\'e}--Cahn derivatives}\label{sc:larchecahnderivaties}

Let
\begin{equation}\label{eq:sets}
\bfvarphi = \{ \varphi^1, \ldots, \varphi^n\}
\end{equation}
be a list of species concentrations and assume that the function $\calF$ depends on $\bfvarphi$ such that
\begin{equation}\label{eq:function1}
\calF(\bfvarphi) = \calF(\varphi^1,\ldots,\varphi^n).
\end{equation}
Constraint~\eqref{eq:const}, with \eqref{eq:concetration.def}, implies that the set of concentrations $\bfvarphi$ must be $0 < \varphi^\alpha < 1$. If we vary one concentration $\varphi^\alpha$ while holding all others fixed violates the constraint~\eqref{eq:const}. Thus, the conventional partial derivative on functions such as $\calF$, on which the constraint~\eqref{eq:const} is active, is not appropriately defined. To overcome this shortcoming, Larch\'e and Cahn~\cite{ larche1978thermochemical} defined the following operation
\begin{equation}\label{eq:lcder}
\frac{\partial^{(\sigma)} \calF(\bfvarphi)}{\partial \varphi^\alpha} = \frac{\text{d}}{\text{d} \epsilon} \calF(\varphi^{1},\ldots,\varphi^\alpha+\epsilon,\ldots,\varphi^\sigma-\epsilon,\ldots,\varphi^{n}) \Bigr|_{\epsilon=0}
\end{equation}
in which we choose any two concentrations $\varphi^\alpha$ and $\varphi^\sigma$ from the set of variables. Then, we introduce an infinitesimal change $\epsilon$ in $\varphi^\alpha$, which induces the opposite infinitesimal variation $\epsilon$ onto $\varphi^\sigma$, while holding all other variables unchanged. Thus, this definition satisfies~\eqref{eq:const} by construction while we express the concentration $\varphi^\sigma$ as
\begin{equation}\label{eq:const1}
\varphi^\sigma = 1 - \sum _{\substack{\alpha=1 \\ \alpha \neq \sigma}} ^{n} \varphi^\alpha.
\end{equation}

In multicomponent Cahn--Hilliard systems, we incorporate cross-diffusion gradient energy coefficients $\Gamma^{\alpha\beta}$ into the free-energy definition and obtain the following free-energy density
\begin{equation}\label{eq:free.energy.form}
\hat{\psi}(\bfvarphi,\Grad\bfvarphi) \coloneqq f(\bfvarphi) + \sum_{\alpha=1}^n\sum_{\beta=1}^n\Gamma^{\alpha\beta}\Grad\varphi^\alpha\cdot\Grad\varphi^\beta.
\end{equation}
Elliott \& Garcke in~\cite{ elliott1997diffusional} prove that multicomponent systems are well-posed when $\Gamma^{\alpha\beta}$  is positive definite, among other conditions. We show that this condition is sufficient but not necessary. To do so, we extend the ideas of Larch{\'e}--Cahn and define a \emph{constrained} inner product on a constrained space. We consider a set of vectors $\{\bfp^\alpha\}$ subject to the following constraint
\begin{equation}\label{eq:p}
\sum_{\alpha=1}^n\bfp^\alpha = \bf0,
\end{equation}
and use the following inner product
\begin{equation}\label{eq:inner.product}
\sum_{\alpha=1}^n\sum_{\beta=1}^n\Gamma^{\alpha\beta}\bfp^\alpha\cdot\bfp^\beta.
\end{equation}

Let each entry of $\Lambda^{\alpha\beta}$ be a single number $\kappa$. Thus, due to~\eqref{eq:p}, $\{\bfp^\alpha\}$ is in the null space of $\Lambda^{\alpha\beta}$, that is, $\text{Null}(\Lambda^{\alpha\beta})=\{\bfp^\alpha\}$. Similarly, if each row of $\Lambda^{\alpha\beta}$ is given by the same entry $\kappa^\beta$, we arrive to the same conclusion. For any of these cases, we have that
\begin{equation}\label{eq:point1}
\sum_{\alpha=1}^n\sum_{\beta=1}^n\Gamma^{\alpha\beta}\bfp^\alpha\cdot\bfp^\beta=\sum_{\alpha=1}^n\sum_{\beta=1}^n(\Gamma^{\alpha\beta}+\Lambda^{\alpha\beta})\bfp^\alpha\cdot\bfp^\beta.
\end{equation}

We impose the constraint~\eqref{eq:p} with respect to the component
$\sigma$ to the quadratic form~\eqref{eq:inner.product} to obtain
\begin{equation}\label{eq:constrained.inner.product}
\sum_{\alpha=1}^n\sum_{\beta=1}^n\Gamma^{\alpha\beta}\bfp^\alpha\cdot\bfp^\beta=\sum_{\substack{\alpha=1 \\ \alpha \neq \sigma}}^{n}\sum_{\substack{\beta=1 \\ \beta \neq \sigma}}^{n}(\underbrace{\Gamma^{\alpha\beta}+\Gamma^{\sigma\sigma}-\Gamma^{\alpha\sigma}-\Gamma^{\sigma\beta}}_{\Gamma^{\alpha\beta}_\sigma}\bfp^\alpha\cdot\bfp^\beta.
\end{equation}
We reinterpret this result as an inner product in an unconstrained space of dimension $n-1$ with a \emph{non-invertible} mapping $\Gamma^{\alpha\beta}\mapsto\Gamma^{\alpha\beta}_{\sigma}$ defined as
\begin{equation}\label{eq:non-invertible}
\Gamma^{\alpha\beta}_{\sigma}\coloneqq\Gamma^{\alpha\beta}+\Gamma^{\sigma\sigma}-\Gamma^{\alpha\sigma}-\Gamma^{\sigma\beta}.
\end{equation}
Consequently, the problem is well-posed if $\Gamma^{\alpha\beta}_{\sigma}$ is positive definite. Moreover, $\Gamma^{\alpha\beta}$ can be indefinite without compromising the well-posedness of the problem. Now, let $\Gamma^{\alpha\beta}$ be a diagonal matrix such that
\begin{equation}\label{eq:diagonal.Lambda}
\Gamma^{\alpha\beta}=\kappa\,\delta^{\alpha\beta}.
\end{equation}
From~\eqref{eq:point1}, we rewrite $\Gamma^{\alpha\beta}$ as
\begin{equation}\label{eq:nondiagonal.Lambda}
\Gamma^{\alpha\beta}=-\kappa(1^{\alpha\beta}-\delta^{\alpha\beta}),
\end{equation}
where $1^{\alpha\beta}$ is a constant matrix populated by ones and $\delta^{\alpha\beta}$ is the Kronecker delta, both of dimension $n$. Although the matrix~\eqref{eq:nondiagonal.Lambda} has a null diagonal, the mapping defined by~\eqref{eq:non-invertible} is identical to the one of the diagonal matrix~\eqref{eq:diagonal.Lambda} for all vectors that satisfy the constraint~\eqref{eq:p}. %To give physical meaning to this inner product, we interpret $\Gamma^{\alpha\beta}\Grad\varphi^\alpha\cdot\Grad\varphi^\beta$ as an interfacial energy between phases $\alpha$ and $\beta$, where a non-zero interaction term of the form $\Gamma^{\alpha\alpha}|\Grad\varphi^\alpha|^2$ is physically meaningless. Additionally, by assuming $\Gamma^{\alpha\beta}$ diagonal, we specify the interfacial interactions of a certain phase with itself as well as that all the interactions between all other phases are identical.

\subsection{Thermodynamics}\label{sc:thermodynamics}

Here, we establish the first and second law of thermodynamics. First, we augment the species balances~\eqref{eq:massbal.pointwise.nosupply}
\begin{equation}\label{eq:massbal.pointwise}
\dot{\varrho}^\alpha=-\Div\bfj^\alpha+s^\alpha+s^\alpha_{\text{ext}},
\end{equation}
to consider an external mass supply $s^\alpha_{\text{ext}}$ as well as an internal one $s^\alpha$ arising from chemical reactions.

We treat chemical reactions in a similar fashion as Gurtin \& Vargas \cite{gurtin1971classical}. Moreover, following Gurtin \cite{gurtin1996generalized} and Cherfils et al. \cite{cherfils2011cahn}, see also \cite{miranville2001consistent,bonfoh2001cahn}, we separate conservation statements from constitutive equations. Thus, we introduce the external power expenditure $\calW_{\text{ext}}$ to $\prt$ done by the external microforces on $\prt$ and microtractions on $\calS$ to describe the thermodynamics of this system as follows
\begin{equation}\label{eq:external.power}
\calW_{\text{ext}}(\prt)\coloneqq\sum_{\alpha=1}^{n}\left\{\int\limits_{\prt}\gamma^\alpha\dot{\varphi}^\alpha\dv+\int\limits_{\calS}\xis^\alpha\dot{\varphi}^\alpha\da\right\},
\end{equation}
where $n$ is the total number of species and $\xis^\alpha=\bfxi^\alpha\cdot\bfn$ is the $\alpha$-th microtraction.

The first law of thermodynamics states the energy balance between the interleaving of internal energy and the expenditure rate of the chemical (diffusion and reaction) power.  The entropy production imbalance, or the second law of thermodynamics in the form of the Clausius-Duhem inequality, states that the rate of growth of the entropy is at least commensurate with the entropy flux and its supply. Thus, we can express these two laws as
\begin{equation}\label{eq:partwise.thermodynamical.laws}
  \left\{\,
    \begin{aligned}
      \dot{\overline{\int\limits_{\prt}\varrho\mskip2mu\varepsilon\dv}}&=\calW_{\text{ext}}(\prt)-\int\limits_{\calS}\bfq\cdot\bfn\da+\int\limits_{\prt}\varrho\mskip2mu r\dv+\sum_{\alpha=1}^{n}\left\{-\int\limits_{\calS}\mu^{\alpha}\bfj^\alpha\cdot\bfn\da+\int\limits_{\prt}\mu^{\alpha}s^\alpha_{\text{ext}}\dv\right\},\\[4pt]
      \dot{\overline{\int\limits_{\prt}\varrho\mskip2mu\eta\dv}}&\ge-\int\limits_{\calS}\frac{\bfq}{\vartheta}\cdot\bfn\da+\int\limits_{\prt}\frac{\varrho\mskip2mu r}{\vartheta}\dv,
    \end{aligned}
  \right.
\end{equation}
where $\varepsilon$ and $\eta$ represent the internal-energy and entropy densities, respectively, $\bfq$~is the heat flux, $r$~is the heat supply, and $\vartheta>0$~is the absolute temperature. There is no contribution of $s^\alpha$ to the energy balance~\eqref{eq:partwise.thermodynamical.laws}.

Using the external power expenditure~\eqref{eq:external.power}, the microforce balance~\eqref{eq:massbal.pointwise}, and the constituent content balance~\eqref{eq:massbal.pointwise}, we can localize the first two laws of thermodynamics~\eqref{eq:partwise.thermodynamical.laws} to
\begin{equation}\label{eq:pointwise.thermodynamical.laws}
\left\{\,
\begin{aligned}
\varrho\mskip2mu\dot\varepsilon&=\sum_{\alpha=1}^{n}\left\{-\pi^\alpha\dot\varphi^\alpha+\bfxi^\alpha\cdot\Grad\dot\varphi^\alpha-\bfj^\alpha\cdot\Grad\mu^{\alpha}+\mu^\alpha(\varrho\mskip2mu\dot{\varphi}-s^\alpha)\right\}-\Div\bfq+\varrho\mskip2mu r,\\[4pt]
\varrho\mskip2mu\dot\eta&\ge-\Div\frac{\bfq}{\vartheta}+\frac{\varrho\mskip2mu r}{\vartheta}.
\end{aligned}
\right.
\end{equation}
Rewriting~\eqref{eq:pointwise.thermodynamical.laws}$_2$, we obtain
\begin{equation}\label{eq:.thermodynamics.2.pointwise}
\varrho\mskip2mu\dot\eta\ge-\frac{1}{\vartheta}\Div\bfq+\frac{1}{\vartheta^2}\bfq\cdot\Grad\vartheta+\frac{\varrho\mskip2mu r}{\vartheta}.
\end{equation}
We now define the free-energy density as
\begin{equation}\label{eq:free.energy.def}
\psi\coloneqq\ep-\vartheta\eta,
\end{equation}
which allow us to rewrite the equation system in terms of $\vartheta$ and $\psi$. To employ this transformation, we multiply~\eqref{eq:.thermodynamics.2.pointwise} by $\vartheta$ and subtract the result from~\eqref{eq:pointwise.thermodynamical.laws}$_1$ to express the pointwise free-energy imbalance as
\begin{equation}\label{eq:pointwise.free.energy.imbalance.conserved}
\varrho\mskip2mu(\dot\psi+\dot{\vartheta}\eta)+\sum_{\alpha=1}^{n}\left\{(\pi^\alpha-\varrho\mskip2mu\mu^{\alpha})\dot\varphi^\alpha-\bfxi^\alpha\cdot\Grad\dot\varphi^\alpha+\bfj^\alpha\cdot\Grad\mu^{\alpha}+\mu^\alpha s^\alpha\right\}+\frac{1}{\vartheta}\bfq\cdot\Grad\vartheta\le0.
\end{equation}

\begin{rmk}[Alternative derivation--Principle of virtual power]
  The definition of virtual power expenditure encompasses internal and external contributions. Internally to $\prt$, the power exerted by internal microforces and the microstresses; while externally to $\prt$, the power effected by the external microforces on $\prt$ and microtractions on $\calS$. This definition assumes that these contributions equilibrate each other, that is,
\begin{equation}\label{eq:virtual.power.principle}
\calV_{\text{int}}(\prt,\chi^\alpha)=\calV_{\text{ext}}(\prt;\chi^\alpha)
\end{equation}
where the definitions of the internal and external virtual powers are
\begin{equation}\label{eq:internal.virtual.power}
\calV_{\text{int}}(\prt;\chi^\alpha)\coloneqq\sum_{\alpha=1}^{n}\left\{\int\limits_{\prt}(-\pi^\alpha\chi^\alpha+\bfxi^\alpha\cdot\Grad\chi^\alpha)\emph{\text{d}}v\right\}
\end{equation}
and
\begin{equation}\label{eq:external.virtual.power}
\calV_{\text{ext}}(\prt;\chi^\alpha)\coloneqq\sum_{\alpha=1}^{n}\left\{\int\limits_{\prt}\gamma^\alpha\chi^\alpha\dv+\int\limits_{\calS}\xis^\alpha\chi^\alpha\emph{\text{d}}a\right\},
\end{equation}
where $\{\chi^\alpha\}$ is a set of $n$ kinematically admissible fields. Finally, we apply the divergence theorem to~\eqref{eq:virtual.power.principle} and use standard variational arguments to localize the balance of microforces~\eqref{eq:massbal.pointwise} to the following. For a more general approach, see \cite{Esp20}.
\end{rmk}

\subsection{Theory of reacting materials}\label{sc:theoryofreactingmaterial}

Theoretically, the total number $m$ of possible independent chemical reactions, where $m \ge n_s\in\bbN$, is not arbitrary. We seek to fit our framework in the thermochemistry theory of reacting materials (see, \cite{truesdell1960classical} and \cite{bowen1968stoichiometry,bowen1968thermochemistry}). Thus, we also postulate the indestructibility of the atomic substances
\begin{equation}\label{eq:indestructibility}
\sum_{\alpha=1}^n \dfrac{t^{\alpha\iota} s^\alpha}{m^\alpha}=0, \qquad 1 \le \iota \le n_a,
\end{equation}
where $n_a\in\bbN$ is the number of atomic substances making up all the components $\calA$, $m^\alpha$ is the molecular weight of the $\alpha$-th component, and $t^{\alpha\iota}$ is a non-negative integer expressing the number of atoms of the $\iota$-th atomic substance present in the $\alpha$-th component. This postulate assumes that the atomic substance are indestructible. Moreover, usually $t^{\alpha\iota}$ is not a square matrix and $\text{rank}(t^{\alpha\iota})=\min(n,n_a)$. Finally, the maximum number of possible chemical reactions is
\begin{equation}\label{eq:independent.chemical.reactions}
m \coloneqq n-\text{rank}(t^{\alpha\iota}).
\end{equation}
In this setting, forward reactions and their reciprocal backward reaction are not independent. Thus, we represent them as a single, effective, chemical reaction.

\subsection{Configurational stress and force}\label{sc:configurationalstresses}

We describe the configurational stress, and the internal and external forces arising in multicomponent systems. We first establish how configurational forces expend power in a migrating control volume $\prt^\prime$. We define $\bfq$ as the migrating boundary velocity acting on $\calS^\prime$ with $\bfn^\prime$ being its outward unit normal. We also refer the reader to \cite{Fri06b,Esp19,Esp17}.

For a migrating volume $\prt^\prime$ the constituent content balance~\eqref{eq:massbal.pointwise} in the partwise form specializes to
\begin{equation}\label{eq:partwise.mass.balance.migrating}
\dot{\overline{\int\limits_{\prt^\prime}\varrho\mskip2mu\varphi^\alpha\dv}}-\int\limits_{\calS^\prime}\varrho\mskip2mu\varphi^\alpha\bfq\cdot\bfn^\prime\da=-\int\limits_{\calS^\prime}\bfj^\alpha\cdot\bfn^\prime\da+\int\limits_{\prt^\prime}s^\alpha\dv.
\end{equation}

We use the external virtual power~\eqref{eq:external.virtual.power}, where $\gamma^\alpha$ and ${\xis^\alpha}$ are conjugate to $\dot{\varphi}^\alpha$. We set as virtual field the advective term $\dot{\varphi}^\alpha+\Grad\varphi^\alpha\cdot\bfq$ to follow the motion of $\calS^\prime$ augmented by the fact that the configurational traction $\bfC\bfn^\prime$ is power conjugate to $\bfq$ on $\calS^\prime$. Since
\begin{equation}
{\xis^\alpha}(\dot\varphi^\alpha+\Grad\varphi^\alpha\cdot\bfq)=(\bfxi^\alpha\cdot\bfn^\prime)\dot\varphi^\alpha+(\Grad\varphi^\alpha\otimes\bfxi^\alpha)\bfn^\prime\cdot\bfq,
\end{equation}
 we arrive at an expression of the total external configurational power
\begin{equation}\label{eq:configurational.power}
\calW_{\text{ext}}(\prt^\prime)=\int\limits_{\calS^\prime}\left(\bfC+\sum_{\alpha=1}^{n}\Grad\varphi^\alpha\otimes\bfxi^\alpha\right)\bfn^\prime\cdot\bfq\da+\sum_{\alpha=1}^{n}\left\{\int\limits_{\prt^\prime}\gamma^\alpha\dot\varphi^\alpha\dv+\int\limits_{\calS^\prime}{\xis^\alpha}\dot\varphi^\alpha\da\right\}.
\end{equation}
The relevant part of the motion of $\calS^\prime$ only involves its normal component $\bfq\cdot\bfn^\prime$. Thus, the power expended is indifferent to the tangential component of $\bfq$, yielding
\begin{equation}\label{eq:configurational.relation}
\bfC+\sum_{\alpha=1}^{n}\Grad\varphi^\alpha\otimes\bfxi^\alpha\eqqcolon\zeta\id,
\end{equation}
where $\zeta$ is a scalar field.

Thus, the first integral of~\eqref{eq:configurational.power} becomes
\begin{equation}
\int\limits_{\calS^\prime}\zeta\mskip+2.5mu\bfq\cdot\bfn^\prime\da.
\end{equation}
The arguments that led to the free-energy imbalance~\eqref{eq:pointwise.free.energy.imbalance.conserved}, allow us to analyze isothermal processes in a migrating control volume $\prt^\prime$ with a velocity $\bfq$. Hence, we arrive at
\begin{multline}
\dot{\overline{\int\limits_{\prt^\prime}\varrho\mskip2mu\psi\dv}}=\int\limits_{\prt^\prime}\varrho\mskip2mu\dot{\psi}\dv+\int\limits_{\calS^\prime}\varrho\left(\psi-\sum_{\alpha=1}^{n}\mu^\alpha\varphi^\alpha\right)\bfq\cdot\bfn^\prime\da\le\\\sum_{\alpha=1}^{n}\left\{\int\limits_{\prt^\prime}\gamma^\alpha\dot\varphi^\alpha\dv+\int\limits_{\calS^\prime}{\xis^\alpha}\dot\varphi\da-\int\limits_{\calS^\prime}\mu^{\alpha}\bfj^\alpha\cdot\bfn^\prime\da+\int\limits_{\prt^\prime}\mu^{\alpha}s^\alpha_{\text{ext}}\dv\right\}+\int\limits_{\calS^\prime}\zeta\bfq\cdot\bfn^\prime\da,
\end{multline}
 leading to
\begin{multline}\label{eq:configurational.free.energy.imbalance}
\int\limits_{\prt^\prime}\varrho\mskip2mu\dot{\psi}\dv\le\sum_{\alpha=1}^{n}\left\{\int\limits_{\prt^\prime}\gamma^\alpha\dot\varphi^\alpha\dv+\int\limits_{\calS^\prime}{\xis^\alpha}\dot\varphi^\alpha\da-\int\limits_{\calS^\prime}\mu^{\alpha}\bfj^\alpha\cdot\bfn^\prime\da+\int\limits_{\prt^\prime}\mu^{\alpha}s^\alpha_{\text{ext}}\dv\right\}\\+\int\limits_{\calS^\prime}\left(\zeta-\varrho\mskip2mu\left(\psi-\sum_{\alpha=1}^{n}\mu^\alpha\varphi^\alpha\right)\right)\bfq\cdot\bfn^\prime\da,
\end{multline}
 which implies that
\begin{equation}\label{eq:zeta}
\zeta\coloneqq\varrho\left(\psi-\sum_{\alpha=1}^{n}\mu^\alpha\varphi^\alpha\right).
\end{equation}

% \bibliographystyle{elsarticle-num}
% \bibliography{bib}

\end{document}